\documentclass[]{interact}

\usepackage{epstopdf}
\usepackage[caption=false]{subfig}

\usepackage[numbers,sort&compress]{natbib}
\bibpunct[, ]{[}{]}{,}{n}{,}{,}
\makeatletter
\def\NAT@def@citea{\def\@citea{\NAT@separator}}
\makeatother

\theoremstyle{plain}

\theoremstyle{definition}

\theoremstyle{remark}

\usepackage{sistyle}
\usepackage{amssymb}
\renewcommand{\vec}{\boldsymbol}

\usepackage[dvipsnames]{xcolor}

\begin{document}

\title{Tornado-Like~Vortices in the Quasi-Cyclostrophic Regime of Coriolis-Centrifugal Convection}

\author{
\name{Susanne Horn\textsuperscript{a,b}\thanks{CONTACT S.~Horn. Email: susanne.horn@coventry.ac.uk} and Jonathan M. Aurnou\textsuperscript{b}}
\affil{\textsuperscript{a}Centre for Fluid and Complex Systems, Coventry University, Coventry CV1 5FB, UK; \textsuperscript{b}Department of Earth, Planetary, and Space Science, University of California, Los Angeles, Los Angeles, CA 90095-1567, USA}
}

\maketitle

\begin{abstract}
Coriolis-centrifugal convection (C$^3$) in a cylindrical domain constitutes an idealised model of tornadic storms, where the rotating cylinder represents the mesocyclone of a supercell thunderstorm. We present a suite of C$^3$ direct numerical simulations, analysing the influence of centrifugal buoyancy on the formation of tornado-like vortices (TLVs). TLVs are self-consistently generated provided the flow is within the quasi-cyclostrophic (QC) regime in which the dominant dynamical balance is between pressure gradient and centrifugal buoyancy forces.
This requires the Froude number to be greater than the radius-to-height aspect ratio, $Fr \gtrsim \gamma$. We show that the TLVs that develop in our C$^3$ simulations share many similar features with realistic tornadoes, such as azimuthal velocity profiles, intensification of the vortex strength, and helicity characteristics. Further, we analyse the influence of the mechanical bottom boundary conditions on the formation of TLVs, finding that a rotating fluid column above a stationary surface does not generate TLVs if centrifugal buoyancy is absent. In contrast, TLVs are generated in the QC regime with any bottom boundary conditions when centrifugal buoyancy is present. 
Our simulations bring forth insights into natural supercell thunderstorm systems by identifying properties that determine whether a mesocyclone becomes tornadic or remains non-tornadic. For tornadoes to exist, a vertical temperature difference must be present that is capable of driving strong convection. Additionally, our $Fr \gtrsim \gamma$ predictions dimensionally imply a critical mesocyclone angular rotation rate of $\widetilde{\Omega}_{mc} \gtrsim \sqrt{g/H_{mc}}$. Taking a typical mesocyclone height of $H_{mc}\approx \SI{12}{km}$, this translates to $\widetilde{\Omega}_{mc}\gtrsim 3~\times~\SI{10^{-2}}{s^{-1}}$ for centrifugal buoyancy-dominated, quasi-cyclostrophic tornadogenesis. The formation of the simulated TLVs happens at all heights on the centrifugal buoyancy time scale $\tau_{cb}$. This implies a roughly 1 minute, height-invariant formation for natural tornadoes, consistent with recent observational estimates.
\end{abstract}


\section{Introduction}
\begin{figure}[ht!]
\includegraphics[width=34pc]{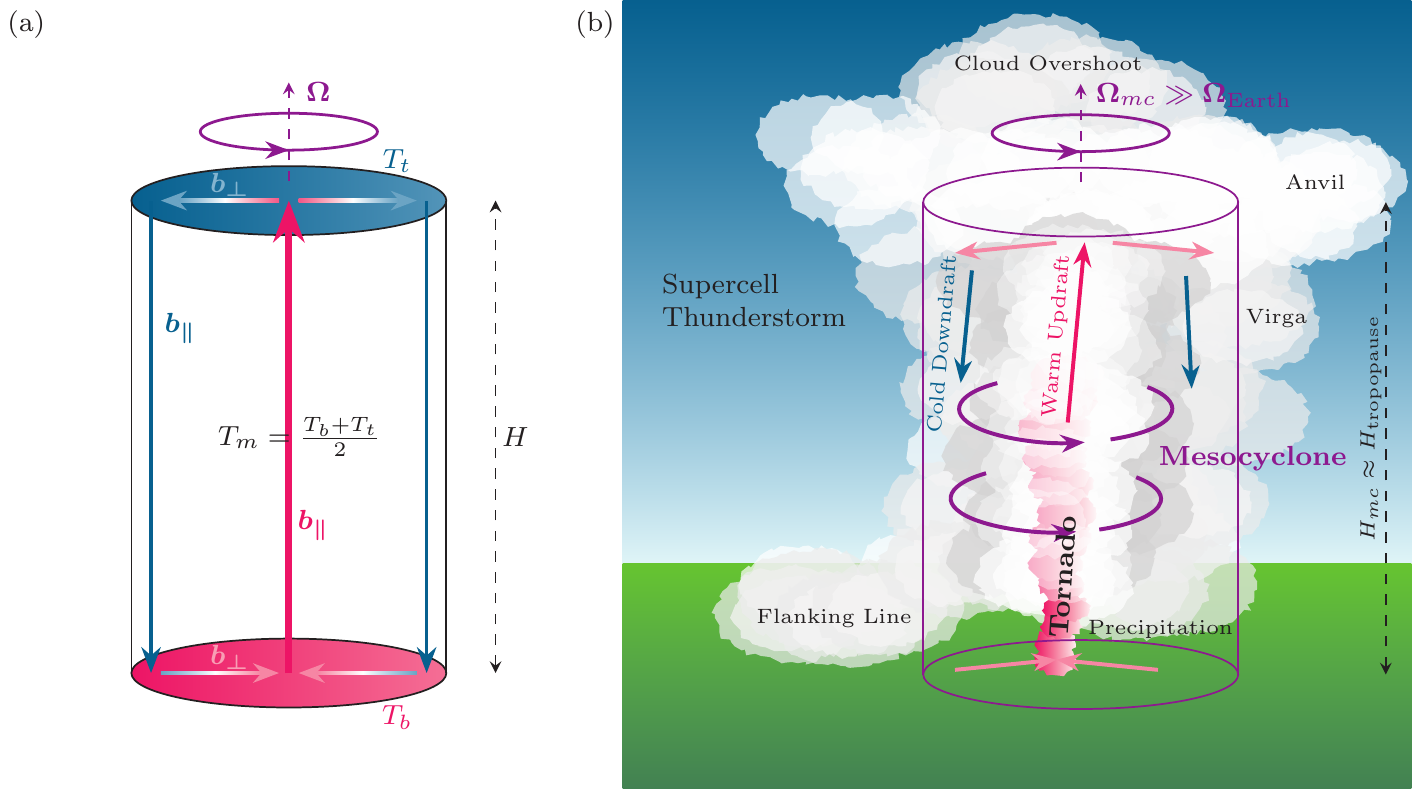}
 \caption{(a) Sketch of the cylindrical rotating Rayleigh-B\'enard convection system studied here, as well as of the centrifugally driven meridional circulation. The non-dimensional top temperature is $T_t = -1/2$ and the bottom temperature is $T_b = 1/2$. The vertical arrows indicate the direction of the flow due to the gravitational buoyancy $\vec{b}_\parallel$ and radial arrows the direction of the flow due the centrifugal buoyancy $\vec{b}_\perp$. Note that if $Fr \neq 0$ the top-bottom anti-symmetry of the system is broken, hence, the arithmetic mean temperature $T_m$ is usually non-zero \citep{Horn2019}. Furthermore, non-linearities and Coriolis effects can lead to significantly altered flow configurations. (b) Schematic of a supercell thunderstorm with a mesocyclone and tornado (not drawn to scale). The cylinder marks the mesocyclonic region rotating with angular velocity $\bm \Omega_{mc}$ that is studied here with idealized direct numerical simulations.
 \label{fig:sketch}}
\end{figure}

The canonical system of rotating Rayleigh-B\'enard convection (RBC) has proved invaluable in elucidating the flow dynamics in many geophysical and astrophysical settings \citep{Horn2018, Horn2019, Aurnou2015, Aurnou2018, Chandrasekhar1961, Cheng2018, Ecke2014, Favier2014, Featherstone2016, Horn2014a, Horn2015, Horn2017, Julien1996c, Julien2012,  Kunnen2008b, Kunnen2013, Kunnen2016, Stellmach2014, Stevens2010b, Stevens2011, Stevens2013, Weiss2010, Weiss2011a, Zhong2009}. The setup consists of a fluid confined between an isothermally heated boundary at the bottom and a cooled boundary at the top that is rotated around the vertical axis as sketched in figure~\ref{fig:sketch}(a). Rotational effects are considered in terms of the Coriolis force alone in the majority of numerical and theoretical studies, since in many natural settings the centrifugal buoyancy term is arguably small even though the centrifugal buoyancy force warrants explicit inclusion within the Oberbeck-Boussinesq approximation \citep{Oberbeck1879, Boussinesq1903, Becker2006, Barcilon1967, Brummell2000,  Curbelo2014, Hart1999, Hart2000, Homsy1969, Homsy1971b, Horn2018, Horn2019, Lopez2006, Lopez2009, Marques2007, Torrest1974}. 

In distinction, the system of rotating Rayleigh-B\'enard convection where the full inertial term is accounted for is referred to as Coriolis-centrifugal convection (C$^3$) \citep{Horn2018, Horn2019}. It has been shown that in C$^3$ vortices that bear a physical and visual resemblance to tornadoes can be obtained. The objective of the present paper is to show that centrifugal buoyancy is relevant for naturally occurring tornadoes and that it connects certain mesocyclone properties with tornadogenesis. Indeed, tornadoes are known to be in a cyclostrophic balance with a dominant force balance between the pressure gradient and centrifugal force. Thus, it is reasonable to assume that centrifugal buoyancy matters for their dynamics. However, so far the importance of centrifugation has only been recognized within the tornado physics community in terms of hydrometeors and debris \citep{Dowell2005, Orf2017}, not in terms of buoyancy, rendering the treatment of centrifugation incomplete.

Tornadoes are the most intense atmospheric vortices. Except for Antarctica, they have been observed on all of Earth's continents \citep{Goliger1998}, as well as on Mars \citep{Balme2006, Thomas1985}. The strongest, most dangerous, and most damaging tornadoes develop in the updraft of supercell thunderstorms. Supercell thunderstorms form when large cold and warm air masses collide and most commonly occur in the midlatitudes, in particular, the Great Plains of the United States. The updraft is formed through wind shear that creates horizontal vorticity which due to the sun heating the ground gets tilted upwards. The combination of warm, humid air rising and strong vertical wind shear leads to a horizontal spin of the updraft. This rotating aircolumn is called mesocyclone, shown schematically in figure~\ref{fig:sketch}(b). Thus, the tornadoes forming in such storms have an associated parent circulation \citep{Davies-Jones2015, Klemp1987, Lemon1979}. Supercell tornadoes are also called type I tornadoes \citep{Bluestein2013}. The vertical vorticity of type I tornadoes is one to two orders of magnitude higher than their parent storm's, and two to three orders of magnitude higher than that of a hurricane or typhoon \citep{Lin1992, Snow1987}. In contrast, tornadoes without a parent circulation, including dust devils, water spouts, and fire whirls, are categorized as type II. Both types are thought to be generated through distinct mechanisms \citep{Bluestein2013, Lemon1979, Maxworthy1973}, with neither mechanism being fully understood. 

In the here considered C$^3$ system, the parent circulation is supplied externally and the rotating cylinder of fluid mimics the mesocylone, as shown in figure~\ref{fig:sketch}. Hence, our model is most relevant for tornadoes type I. 
The existence of an intimate relationship between tornadoes type I and their harboring supercell mesocyclones is well established, but one of the unsolved mysteries of tornado research revolves around the question what makes a given mesocyclone tornadic \citep{Wurman2012}. In fact, less than 25\% of all mesocyclones spawn tornadoes, and mesocyclones with arguably similar properties may or may not generate tornadoes \citep{Markowski2014, Trapp2005}. Further, it is not known what is required to maintain a tornado, nor what ultimately leads to its demise \citep{Marquis2012}. Similarly, it is also not yet possible to predict a tornado's intensity or duration \citep{Markowski2014}. 
Hence, identifying the supercell mesocyclone characteristics that allow researchers to answer these questions remains a major challenge. The current inability to do so indicates that least one crucial physical mechanism is still missing \citep{Bluestein2013, Trapp2005}. The three main approaches that aim to tackle this challenge are observational field campaigns \citep{Rasmussen1994, Wurman2012}, simulations of the entire supercell thunderstorm \citep{Orf2017}, and idealized local laboratory and numerical models of tornado-like vortices (TLVs) \citep{Fiedler1995, Ward1972}.

The observational approach is the most direct one, but it comes with difficulties, first and foremost, having the measurement equipment at the right place at the right time. Second, tornadoes are often problematic to sense visually. They require either tracer particles such as dust or a low enough pressure that leads to condensation in order for them to be optically detectable \citep{Bluestein2013}. Thus, frequently only a condensation funnel is seen aloft despite the tornado existing at the ground. Third, tornadoes and their generation are characterized by short time scales, being of order of only a couple seconds to a bit more than an hour with an average of ten minutes. These time scales are shorter than the volumetric update times of older Doppler radar scans \citep{Houser2018, Rotunno2013}.

The latter two issues have led to the long-held belief that the majority of tornadoes ``touch down'', i.e., build down from aloft via the dynamic pipe effect \citep{Leslie1971, Smith1978, Trapp1997}. The dynamic pipe effect assumes that there is initially a cyclostrophic vortex at a certain height. In a cyclostrophic balance, radial inflow into the vortex is forbidden, because per definition, the radial pressure gradient force equals the centrifugal force. Vertical inflow, on the other hand, is permitted by the vertical pressure gradient force. Thus, the vortex can act as a pipe by sucking in air, thereby concentrating vorticity at the lower end of the vortex. At this lower level, a new cyclostrophic balance establishes itself. The process continues, the vortex grows further towards the ground and simultaneously shrinks in radius until it ultimately touches down.

Recent measurements do not support this scenario and suggest a paradigm shift. Rapid Doppler radar scans with fast volumetric update times of up to ten seconds proved that the majority of tornadoes form on time scales of about one minute, hence, much shorter than the several tens of minutes required by the dynamic pipe effect \citep{Trapp1999, Houser2015}. Most tornadoes are now believed to build from the bottom up or are formed almost simultaneously along their entire vertical axis \citep{Houser2015, French2013, Bluestein2019, Houser2018}. Recognizing the dynamic pipe effect as nonessential, or possibly obsolete, makes the search for the correct physical mechanisms even more pressing.

Valuable additional insights that complement radar measurements are provided by numerical simulations of entire supercell thunderstorms. These type of numerical simulation utilise field measurements for initialisation and are case studies of specific storm events \citep{Orf2017}. But with a minimum possible grid spacing of \SI{30}{m} compared to an average tornado scale of \SI{100}{m}, small scale processes within the tornado remain unresolved and require the need of turbulence and microphysics modeling. Further, the full tornado parameter space cannot be explored and generalisations are difficult.

The third approach of idealized numerical and laboratory models of TLVs, which is also followed here, remains a promising and useful tool for understanding the fundamental turbulent fluid dynamics of tornadoes.
The common underlying assumption is that only the mesocyclone, i.e. a rotating and, generally, precipitation-free updraft, is needed which is modelled as a rotating column of fluid extending from the ground up to the tropopause \citep{Rotunno2013}, as schematized in figure~\ref{fig:sketch}(b). This approach does not yield insight into the  process of generating the mesocyclone, though, which is thought to involve a first cyclone forming aloft in a barotropic process through tilting of streamwise vorticity and a second low-level, near-ground cyclone generated through baroclinicity that moves underneath \citep{Davies-Jones2015}. Any translatory motion of the storm assumed to be with velocities between 0 and \SI{25}{m/s} \citep{Davies-Jones2015} is also neglected. 

The prototypes of these simplified models are the laboratory Ward chamber and the numerical Fiedler chamber \citep{Fiedler1995, Rotunno2013, Ward1972}, where, in both cases, the geometry is simplified to a cylinder. In the Ward chamber, the updraft is obtained using mechanical forcing through a fan, and angular momentum is supplied by a rotating screen. In the Fiedler chamber, the incompressible Navier-Stokes equations are solved including the Coriolis force and a prescribed radially and vertically dependent, but time-independent, buoyancy force. No temperature equation is considered, i.e., the system is isothermal and no feedback exists between forcing and the generated flows. In addition, a viscosity that artificially increases with height is required for numerical stability, and occasionally LES-style turbulence models are employed \citep[e.g.][]{Lewellen2000}. Both the laboratory and the numerical set-ups successfully produce TLVs, and the original as well as the successor models have substantially improved our understanding of tornadoes \citep{Church1979, Fiedler1994, Fiedler1995, Fiedler1998, Fiedler2009,  Nolan1999, Nolan2012, Nolan2005, Rotunno2013,  Rotunno2016, Vogt2013, Castano2017}.  
\section{Governing Equations and Numerical Methodology}
A natural extension of the Fiedler/Ward model is to explicitly consider the fact that tornadoes are born in a buoyancy-driven convective environment, such that there is a coupling between the temperature and the velocity fields. The rotating updraft within a supercell thunderstorm is idealized as a fluid in a cylinder rotating around its vertical axis with an angular speed $\vec{\Omega} = \Omega \hat{\vec{e}}_z$. Convective energy is made available by imposing a constant adverse vertical temperature difference $\Delta = T_b - T_t$ between the bottom and the top, see figure~\ref{fig:sketch}.

Since compressibility effects are argued to be small for tornadoes \citep[e.g.][]{Lewellen1993, Rotunno2016}, we employ the Oberbeck-Boussinesq (OB) approximation that leads to a solenoidal velocity field, i.e., an incompressible continuity equation. This constitutes a significant simplification, but the advantage of alleviated numerical costs preponderate, especially, as typically the OB approximation yields accurate results well outside of its formal range of validity  \citep{Ahlers2006, Gray1976, Horn2013a, Horn2014a, Sugiyama2009}. Moreover, in the present context, the OB approximation also has the advantage that it allows us to explicitly isolate the effect of centrifugal buoyancy. 
That is, the gravitational buoyancy force appears in the vertical momentum equation, and the centrifugal buoyancy force in the radial momentum equation,
\begin{equation}
\vec{b}_\parallel = g \alpha (T -T_m) \hat{\vec{e}}_z, \quad \vec{b}_\perp =  -\Omega^2 r \alpha (T-T_m) \hat{\vec{e}}_r,
\end{equation}
respectively \cite{Marques2007, Horn2019}. Here, $g$ is the gravitational acceleration, $\alpha$ the isobaric expansion coefficient, and $T_m = (T_b +T_t)/2$ the reference and arithmetic mean temperature. The parallel symbol $\parallel$ indicates that the gravitational acceleration $g \hat{\vec{e}}_z$ is parallel to the imposed temperature gradient and rotation vector, and the perpendicular symbol $\perp$ indicates that the centrifugal acceleration $\Omega^2 r \hat{\vec{e}}_r$ is perpendicular to the imposed temperature gradient and rotation vector.

The usually neglected centrifugal buoyancy force $\vec{b}_\perp$ and the more familiar gravitational buoyancy force $\vec{b}_\parallel$ act in similar ways: $\vec{b}_\parallel$ results in warm ($T > T_m$) and, thus, less dense fluid parcels moving upwards and the cold ($T < T_m$) and, thus, denser fluid parcels moving downwards. Similarly, $\vec{b}_\perp$ results in warmer, less dense fluid moving radially towards the centre and colder, denser fluid moving radially away from the centre. Hence, simplistically, one may expect that both forces induce a meridional circulation as sketched in figure~\ref{fig:sketch}(a).

The governing set of equations for the velocity field $\vec{u}$ and the temperature $T$ are the incompressible continuity, the Navier-Stokes and the temperature equation. They read
\begin{eqnarray}
\vec{\nabla} \cdot \vec{u} &=& 0, \label{eq:NS1}\\
D_t \vec{u} &=& \nu \vec{\nabla}^2 \vec{u} - \vec{\nabla}p  + 2 \Omega  \vec{u}  \times \hat{\vec{e}}_z  - \Omega^2 r \alpha (T-T_m) \hat{\vec{e}}_r + g \alpha (T -T_m) \hat{\vec{e}}_z,\label{eq:NS2}\\
D_t T &=&  \kappa \vec{\nabla}^2 {T},\label{eq:NS3}
\end{eqnarray}
where $\nu$ denotes the kinematic viscosity and $\kappa$ the thermal diffusivity both evaluated at $T = T_m$.

The eqs.~\eqref{eq:NS1}--\eqref{eq:NS3} can be non-dimensionalized by introducing appropriate reference scales, which allows for the interpretation of the solutions in a more general fashion compared to the dimensional equations. Here, the non-dimensional temperature is given by $\breve{T} = (T - T_m)/\Delta$, the non-dimensional spatial coordinates by $\breve{\vec{x}} =  \vec{x}/R$, and the non-dimensional velocity by $\breve{\vec{u}} = \vec{u}/\sqrt{\alpha \Delta g  R}$, with $R$ being the radius of the cylindrical domain. Accordingly, the non-dimensional time is given by $\breve{t} = t R/ \sqrt{ \alpha \Delta g R}$, and the non-dimensional pressure by $\breve{p} = p/(\rho \alpha \Delta g R)$. With this choice of reference scales, the following non-dimensional set of equations is obtained:
\begin{eqnarray}
 \vec{\nabla} \cdot \breve{\vec{u}} &=& 0, \label{eq:NSnd1}\\
D_{\breve{t}} \breve{\vec{u}} &=&  \displaystyle  \sqrt{ \frac{Pr}{ Ra \, \gamma^3}} \, \vec{\nabla}^2 \breve{\vec{u}} - \vec{\nabla} \breve{p}  + \sqrt{\frac{\gamma}{ Ro_\parallel^2}}   \breve{\vec{u}} \times \hat{\vec{e}}_z - Fr \, \breve{T} \, \breve{r} \, \hat{\vec{e}}_r +  \breve{T} \hat{\vec{e}}_z,\label{eq:NSnd2}\\
D_{\breve{t}} \breve{T} &=&   \displaystyle  \sqrt{\frac{1}{ Ra \, Pr\, \gamma^3}}  \vec{\nabla}^2 \breve{T}. \label{eq:NSnd3}
\end{eqnarray}
For clarity the breve marking non-dimensional quantities will be omitted in the following. The prefactors are expressed through the non-dimensional control parameters of the Coriolis-centrifugal convection system \citep{Horn2018, Horn2019}, the Prandtl, Rayleigh, gravitational Rossby, Froude number, and the radius-to-height aspect ratio:
\begin{align}
Pr = \frac{\nu}{\kappa},\quad Ra = \frac{\alpha g \Delta H^3}{\kappa \nu},\quad Ro_\parallel =  \frac{\sqrt{\alpha  g \Delta H}}{2 \Omega H}, \quad Fr  = \frac{\Omega^2 R}{g},\quad \gamma = \frac{R}{H}. 
\end{align}

\begin{figure}[t!]
\centerline{\includegraphics[width=\textwidth]{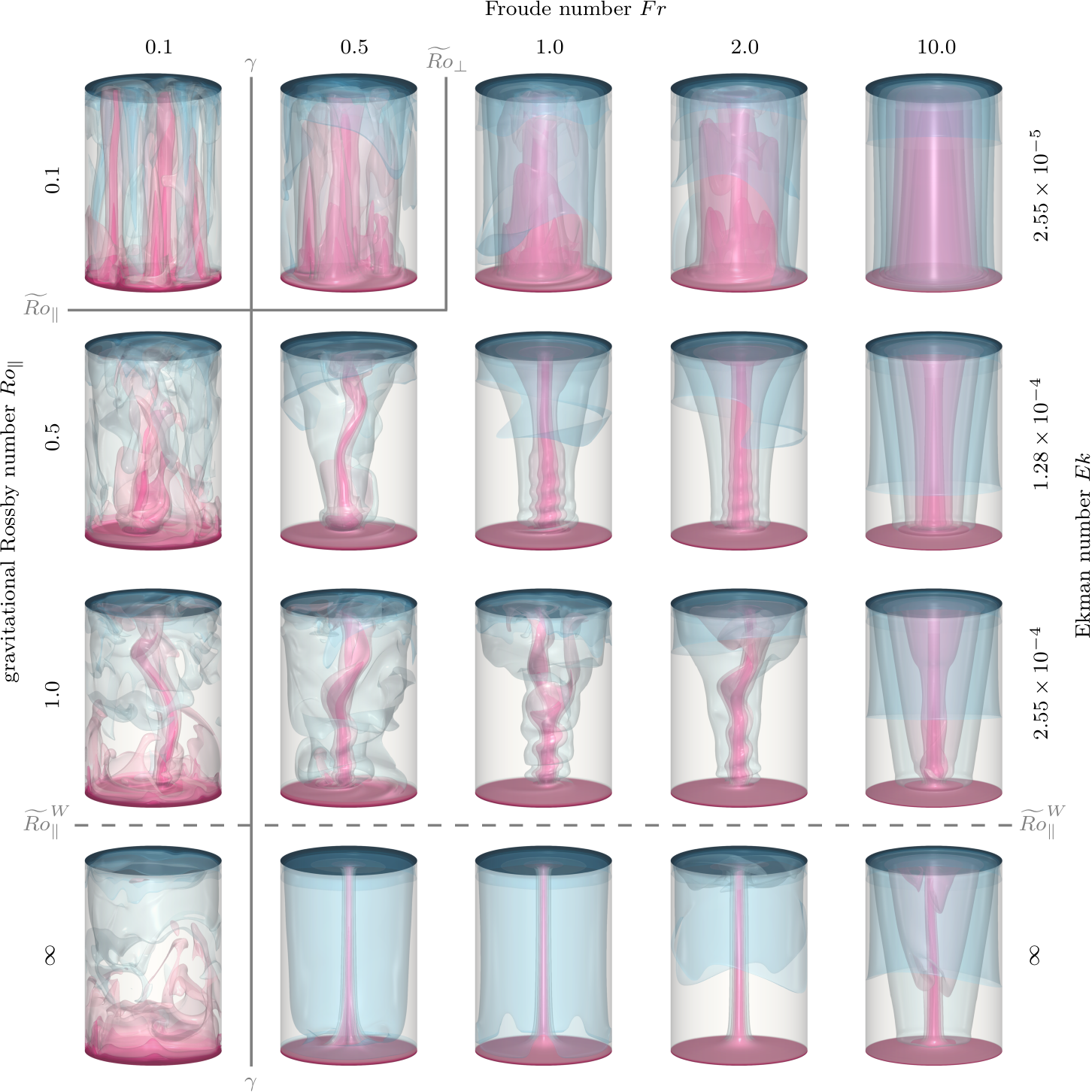}}
\caption{Instantaneous temperature isosurfaces of solutions in the quasi-cyclostrophic (QC) regime of Coriolis-centrifugal convection for fixed $Ra = 10^8$, $Pr = 6.52$, and $\gamma = 0.365$ as function of the gravitational Rossby number $Ro_\parallel$ and the Froude number $Fr$. The Coriolis force increases from bottom to top, and the centrifugal buoyancy force from left to right; blue corresponds to $T < T_m$ and pink to $T>T_m$. The grey lines mark the transitions to the 3D, QG, and CC regime, respectively, given by $\widetilde{Ro}_\parallel$, $\widetilde{Ro}_\perp$, and $\widetilde{Fr} \simeq \gamma$, according to eqs.~\eqref{eq:Ropar}--\eqref{eq:gamma}. The solutions $(Ro_\parallel, Fr)$ = $(\infty, 0.1), \, (1.0, 0.1),\,(0.5,0.1)$ lie in the 3D regime,  $(Ro_\parallel, Fr)$ = $(0.1, 0.1)$ in the QG regime, and $(Ro_\parallel, Fr)$ = $(0.1,0.5)$ in the CC regime.
\label{fig:tornadoes}}
\end{figure}

The eqs.~\eqref{eq:NSnd1}--\eqref{eq:NSnd3} are solved numerically using the well-established finite volume code \textsc{goldfish} \citep{Horn2018, Shishkina2015, Shishkina2016}. We perform direct numerical simulations (DNS) and do not prescribe any turbulence modeling. The simulated parameter space is given by $Ra \in \{10^7, \, 10^8, \, 10^9\}; \; Pr = 6.52, \;0.0125 \leq Ro_\parallel \leq \infty;\;  0 \leq Fr \leq 10;$ and $\gamma \in \{0.365,\, 1.5\}$, with the most comprehensive $Ro_\parallel$ and $Fr$ coverage for $Ra = 10^7$ and $10^8$ and $\gamma = 0.365$. A few single specific cases were conducted for $\gamma = 1.5$, $Pr = 0.7$. The DNS of rotating Coriolis-centrifugal convection, $Ro_\parallel < \infty$ and $Fr \neq 0$, have all been initialized with a non-rotating $Ro_\parallel = \infty$ and $Fr = 0$ flow field. A total number of 165 DNS were performed, and the majority of them have been analysed in a different context before \cite{Horn2018, Horn2019}. The resolutions of the main DNS are $N_r \times N_\phi \times N_z = 180 \times 512 \times 480$ volume cells for $Ra = 10^9$; $142 \times 512 \times 384$ for $Ra = 10^8$; and $47 \times 64 \times 120$ with a few $47 \times 64 \times 130$ cases for $Ra = 10^7$. 
The resolution was chosen following the resolution criteria developed by Shishkina et al. \cite{Shishkina2010}. We took particular care in resolving the Ekman boundary layer with the same number of volume cells as a viscous boundary layer. Since both Coriolis and centrifugal buoyancy generally suppress turbulence, the bulk resolution is less restrictive than in the non-rotating case which was also verified via a few higher-resolved test simulations. An overview of the tornado-like solutions for $Ra = 10^8$ is given in figure~\ref{fig:tornadoes}.
The eqs.~\eqref{eq:NSnd1}--\eqref{eq:NSnd3}  are completed by no-slip conditions at all walls, $\vec{u}|_{wall} = \vec{0}$, isothermal top and bottom, $T_t = -1/2$ and $T_b = 1/2$, and thermally insulated sidewall boundary conditions, $\partial_r T|_{r=R} = 0$. In addition, a few DNS with different mechanical BCs are conducted, as discussed in \textsection~\ref{sec:bcs}.

Our idealised DNS cannot match all of the complexities and facets of the natural system. Several effects are not included in order to focus on the implications of centrifugal buoyancy on the flow, as well as to lower the numerical costs. Effects that have been ignored include compressibility, condensation and evaporation, hydrometeors, dust, surface roughness, vibration-induced boundary layer destabilization, and translatory storm motions \citep{Bluestein2013, Davies-Jones2003,  Davies-Jones2015,Doswell2004,  Rotunno2013, Wang2020}. Also, the rotating updraft is unlikely to be a perfectly right cylinder, and along the sidewall and the top boundaries gas exchange is possible as opposed to the impenetrable boundaries employed here.
Due to prohibitive resolution requirements, present day DNS are not able to reach realistic atmospheric values of the Rayleigh number $Ra$, or, equivalently of the Reynolds number $Re$, and, connected to this, of the Ekman number $Ek$. The main effects are a lower level of turbulence, and presumably structures with larger characteristic length scales in the DNS compared to nature. Other control parameters, in particular, $Ro_\parallel$, $Fr$, $\gamma$, $Ro_\perp$, and $\chi$, can be matched and are essential to gain insight into tornadic systems.

\subsection{Non-dimensional parameter space in Coriolis-centrifugal convection \label{sec:c3}}
For the sake of completeness and comparison to natural tornadic systems, we  briefly summarise the C$^3$ control parameters and regime boundary predictions \cite{Horn2018, Horn2019}, and also relate them to non-dimensional parameters found more commonly in the tornado physics literature \cite{Nolan1999, Nolan2005, Nolan2012, Rotunno2013}.
The non-dimensional parameters that appear in the eqs.~\eqref{eq:NSnd1}--\eqref{eq:NSnd3} are ratios of characteristic time scales which help to identify the governing physics \citep[e.g.][]{Cheng2016, Horn2018, Horn2019, Aurnou2020}. The relevant time scales in C$^3$ \citep{Horn2018, Horn2019} are the thermal diffusion time scale $\tau_\kappa = H^2/\kappa$; the viscous diffusion time scale $\tau_\nu = H^2/\nu$; the Coriolis time scale $\tau_\Omega = 1/(2 \Omega)$; the gravitational buoyancy or free-fall time scale $\tau_{f\!f} = H/\sqrt{\alpha \Delta g H}$; and the centrifugal buoyancy time scale $\tau_{cb} = R/ \sqrt{\alpha \Delta \Omega^2 R^2} = 1/\sqrt{\alpha \Delta \Omega^2}$.

The Prandtl number is given by the ratio of the thermal to the viscous diffusion time scale, $Pr = \tau_\kappa/\tau_\nu$, and,  thus, describes whether momentum or heat is transported more efficiently. $Pr$ is a pure material property. For air $Pr \approx 0.7$ and for water $Pr \approx 6.52$. We found that both fluids show qualitatively similar behavior \citep{Horn2014a, Horn2015, Horn2018, Stevens2010b}, but for $Pr = 6.52$ the more coherent temperature field slightly improves the visualization of the flow structures. The Rayleigh number compares the free-fall time scale to both the thermal and the viscous diffusion time scale, $Ra = \tau_\kappa \tau_\nu/\tau_{f\!f}^2$, and is a measure of the thermal forcing. Furthermore, $Ra$ and $Pr$ define a Reynolds number
\begin{equation}
Re = \frac{\sqrt{\alpha \Delta g H} H}{\nu} =   \sqrt{\frac{Ra}{Pr}} 
\end{equation}
where the characteristic velocity is assumed to be the free-fall velocity $\sqrt{\alpha \Delta g H}$, which is identical to the thermodynamic wind speed limit for this set-up \citep{Rotunno2013}. Generally, $Ra$ determines the turbulence level in convective flows. The higher $Ra$ the smaller small-scale structures and vortices become, notwithstanding that the large-scale structures remain of comparable size \citep{Ahlers2009, Grossmann2000}.
The gravitational Rossby number characterizes the importance of rotation in turbulent convection by comparing the Coriolis time scale to the free-fall time scale, $Ro_\parallel = \tau_\Omega/\tau_{f\!f}$. For low Rossby numbers, $Ro_\parallel \ll 1$, flows are in the quasi-geostrophic regime \citep{Julien2012b} whereas $Ro_\parallel = \infty$ corresponds to zero Coriolis force. In tornado vortex chambers the swirl ratio $S$ is used \citep{Nolan2005, Rotunno2013}, which is half of the inverse gravitational Rossby number, i.e.,
\begin{equation}
S = \frac{\Omega H}{\sqrt{\alpha \Delta g H}} =  \frac{1}{2 Ro_\parallel}.
\end{equation}
An additional non-dimensional parameter is defined by the ratio of the Coriolis time scale to the viscous diffusion time scale, $ \tau_\Omega/\tau_\nu$. This ratio defines the Ekman number
\begin{equation} 
Ek = \frac{\nu}{2\Omega H^2} = \sqrt{\frac{Ro_\parallel^2 Pr}{Ra}} = \frac{Ro_\parallel}{Re},
\end{equation}
which is small in quasi-geostrophic flows. Thus, $Ek$ and $Ro_\parallel$ are connected, and they are both important to characterize rotating convection dynamics. We note that in the tornado literature \citep[e.g.][]{Nolan1999, Nolan2005, Nolan2012} half the inverse Ekman number is known as the vortex Reynolds number
\begin{equation}
Re_v = \frac{\Omega H^2}{\nu}  = \frac{1}{2 Ek}.
\end{equation}

In the C$^3$ system, the centrifugal buoyancy time scale $\tau_{cb}$ is also relevant. The ratio $\tau_\Omega/\tau_{cb}$  defines the centrifugal Rossby number
\begin{equation}
Ro_\perp = \frac{\sqrt{\alpha \Delta}}{2} = \sqrt{\frac{Ro_\parallel^2 Fr}{\gamma}}, \label{eq:roperp}
\end{equation}
analogous to the gravitational one; $\sqrt{Ro_\perp}$ is also called the density deficit parameter or the thermal Rossby number \citep{ Barcilon1967, Brummell2000, Hart1999, Hart2000, Homsy1969}. In addition, $\tau_{cb}$ can be used to rewrite the Froude number as $Fr = \gamma \tau_{f\!f}^2 / \tau_{cb}^2 $. Importantly, no related control parameters exist for $Fr$ or $Ro_\perp$ in the tornado physics literature since centrifugal buoyancy has not been explicitly considered. 

The control parameters of the C$^3$ system have been used to quantify the transitions between regimes with different dominating flow physics \citep{Horn2018, Horn2019}. It is assumed that $Ra$ is sufficiently high such that the flow is strongly supercritical, i.e., far away from the onset of convection, and $Pr \sim {\cal O}(1)$. For fixed $Ra$ and $Pr$, the main parameters are, thus, $Ro_\parallel$ and $Fr$. We also assume that Coriolis effects are not completely negligible, that is $Ro_\parallel$ is sufficiently low according to the empirical relation by Weiss et al. \cite{Weiss2010, Weiss2011a, Horn2019},
\begin{equation}
 \widetilde{Ro}{}_\parallel^W \, \lesssim  \, \frac{2\gamma}{a} \left(1+ \frac{b}{2\gamma} \right)^{-1}, \quad a = 0.381, \, b = 0.061. \label{eq:Weiss}
\end{equation}
For $\gamma = 0.365$, Eq.~\eqref{eq:Weiss} corresponds to $\widetilde{Ro}{}_\parallel^W \simeq 1.77$. 

The four main regimes previously identified and discussed in detail by Horn \& Aurnou \cite{Horn2018, Horn2019} are the fully three-dimensional (3D), the quasi-geostrophic (QG), quasi-cyclostrophic (QC), and the Coriolis-centrifugal (CC) regime. Based on time scale arguments the transitions between these regimes were identified \cite{Horn2018, Horn2019, King2012b} as
\begin{eqnarray}
\mbox{3D}\leftrightarrows \mbox{QG}:&& \widetilde{Ro}_\parallel \simeq 5.5\, Pr^{-1/2} Ra^{-1/6},\label{eq:Ropar}\\
\mbox{QG}\leftrightarrows \mbox{CC}:&& \widetilde{Ro}_\perp \simeq 5.5\, Pr^{-1/2} Ra^{-1/6},\label{eq:Roperp}\\
\mbox{3D}\leftrightarrows \mbox{QC},\, \mbox{QG}\leftrightarrows \mbox{CC}:&& \widetilde{Fr} \simeq \gamma.\label{eq:gamma}
\end{eqnarray}
Note that the transitions $\widetilde{Ro}_\parallel$ and $\widetilde{Ro}_\perp$ occur at successively lower values with increasing Rayleigh number. In later sections, we will also consider the more conservative possibility \cite[cf.][]{Stevens2009, Horn2015, Zhong2009} that $\widetilde{Ro}_\parallel$ and  $\widetilde{Ro}_\parallel$ remain fixed and close to the current estimate of $0.1$.

The 3D and QG regimes, where centrifugal buoyancy has negligible influence on the flow dynamics, as well as the transition $\widetilde{Ro}_\parallel$ have been extensively studied \citep{Ecke2014, Horn2014a, Julien2012, Kunnen2011}.
However, the CC and QC regimes, where centrifugal buoyancy $\vec{b}_\perp$ dominates over gravitational buoyancy $\vec{b}_\parallel$, are still relatively unexplored. Eq.~\eqref{eq:gamma} also defines the superfroudeality of a given flow \cite{Horn2019}, 
\begin{equation}
\chi \equiv \frac{Fr}{\widetilde{Fr}} =  \frac{Fr}{\gamma} = \frac{\tau^2_{f\!f}}{\tau^2_{cb}}.\label{eq:super}
\end{equation}
Simply put, when $\chi \gtrsim 1$, $\vec{b}_\perp$ overcomes $\vec{b}_\parallel$. In the CC regime, the flow is quasi-geostrophic, such that the pressure gradient and the Coriolis force are nearly in balance, but where the centrifugal buoyancy force also comes into play at first order. Thus, the dynamically important timescales are ordered as $\tau_\Omega \ll \tau_{cb} \ll \tau_{ff}$. This is also known as gradient wind balance \cite{Willoughby1990}, and describes leading order hurricane and typhoon dynamics. 

The focus of the present paper is the QC regime. Horn \& Aurnou \cite{Horn2018} have shown that TLVs are self-consistently generated in idealised DNS in this regime. The flow is predominantly balanced by the centrifugal buoyancy and the pressure gradient forces, a state known as quasi-cyclostrophic balance. This implies the following ordering of the timescales, $\tau_{cb} \ll \tau_{\Omega}$, $\tau_{cb} \ll \tau_{ff}$.

\section{Tornado-Like~Vortices (TLVs) in the Quasi-Cyclostrophic (QC) Regime \label{sec:QC}}
All our solutions are obtained in a statistically steady state, whereas real tornadoes are transient phenomena.  However, for the cases that develop tornado-like vortices (TLVs), the vortex formation happened almost instantaneously along the entire axis and usually within $t \approx {\mathcal{O}}(\tau_{cb})$. Hereafter, we discuss the qualitative similarities of the TLVs with naturally occurring tornadoes. In particular, we follow Ward \cite{Ward1972}, who listed three essential tornado features any model should be able to reproduce:
\begin{enumerate}
 \item the characteristic pressure profiles, with high surface pressure rings surrounding a low pressure core,
 \item bulging deformations along the vortex axis,
 \item the possible development of multiple vortices, that may have a single centre of convergence.
\end{enumerate}
We then discuss in more quantitative detail the azimuthal velocity profiles, vortex intensification, helicity, and the influence of the bottom boundary conditions.

\subsection{Overview of the Flow Morphologies}

An overview of the flow structures in the quasi-cyclostrophic (QC) regime and the bounding transitional regions for fixed $Ra = 10^8$ is presented in figure~\ref{fig:tornadoes}. 
The visualised temperature fields exhibit a rich variety of flow morphologies, including cone, wedge, hourglass, needle, and rope shapes, similar to those observed in natural tornadoes \citep{Bluestein2013}. 
Due to the relatively low $Ra$ and higher $Pr$, the generated TLVs are very coherent and likely possess a larger radius relative to the cylindrical domain when compared to the natural system of a single tornado within a mesocyclone \citep{Rotunno2013}. 

But not all flows in figure~\ref{fig:tornadoes} exhibit tornadic behaviour, which indicates how sensitive the generation of TLVs is to the two major control parameters $Ro_\parallel$ and $Fr$. Without Coriolis force, i.e. $Ro_\parallel = \infty$, the lowermost row in figure~\ref{fig:tornadoes}, the flow is three-dimensional for $Fr \lesssim \gamma$. For $Fr \gtrsim  \gamma$ a radially converging flow along the bottom plate culminates in a warm, fast central upflow that impinges on the cold top plate. There it diverges, leading to a cold, slower downwelling along the sidewall, much like the schematized meridional circulation shown in figure~\ref{fig:sketch} (a).

A setup with zero Coriolis force and non-zero centrifugal buoyancy force is admittedly only perfectly realisable in numerical simulations as both $Ro_\parallel$ and $Fr$ depend on $\Omega$. Hence, a closer resemblance to genuine tornadoes is obtained in the cases with $Ro_\parallel < \infty$. For rather weak Coriolis forces, $Ro_\parallel = 1.0$ and $Ro_\parallel = 0.5$, the two middle rows in figure~\ref{fig:tornadoes}, an inward spiralling flow along the bottom plate leads to archetypal TLVs. These TLVs inherit the sense of rotation from the imposed rotation rate just as natural tornadoes inherit their sense of rotation from their parent storm.

The TLVs are produced relatively centrally. However, most TLVs orbit irregularly or quasi-periodically around the centre. They usually do not reside exactly at $r = 0$, except for the nearly steady cases.
This is in basic agreement with observations: Tornadoes are commonly formed within the rotating updraft, but not in its very centre. In the rare cases when the tornado is located in the very centre, it can be extraordinarily long-lived \citep{Davies-Jones2015}. Moreover, if an additional translatory movement would have been considered, these quasi-periodic orbital motions would leave epicyclic patterns much like the ones known from observations where the entire storm typically moves with velocities between \SI{0}{m/s} and \SI{25}{m/s} \citep{Bluestein2013}.

For strong Coriolis forces, $Ro_\parallel \lesssim 0.1$ (the uppermost row in figure~\ref{fig:tornadoes}), the centrifugal buoyancy force needs to be stronger for TLVs to develop. Hence, $Fr$ needs to be sufficiently high such that $Ro_\perp \gtrsim \widetilde{Ro}_\perp$. For smaller $Ro_\perp$, the flow is instead in the non-tornadic QC and CC regimes. Hence, in accordance with observations, no tornadoes form if the Coriolis force starts to dominate the dynamics \citep{Davies-Jones1973}. For $Ro_\parallel = 0.1$ and $Fr =  0.1$ only convective Taylor columns characteristic for quasi-geostrophic rotating convection form \citep{Cheng2015, Horn2014a, Sprague2006, Stellmach2014, Grooms2010}. For $Fr = 10.0$ the stabilizing effect of centrifugation and Coriolis force leads to a fully steady solution. The flow itself is characterized by ring structures and, thus, is rather hurricane- or typhoon-like \citep{Horn2018}. However, for the intermediate $0.5 \lesssim Fr \lesssim 2.0$ range structures develop that may still be considered tornado-like. For example, for $Fr = 0.5$ multiple vortices with a single centre of convergence develop, resembling tornado outbreaks where in extreme cases over 30 tornadoes can be observed \citep{Trapp2005, Ward1972}. For $Fr = 1.0$ and $Fr = 2.0$ the central vortex is surrounded by sheaths often found around tornadoes \citep[cf.][]{Snow1984}. 

\begin{figure*}
\centerline{\includegraphics[width=\textwidth]{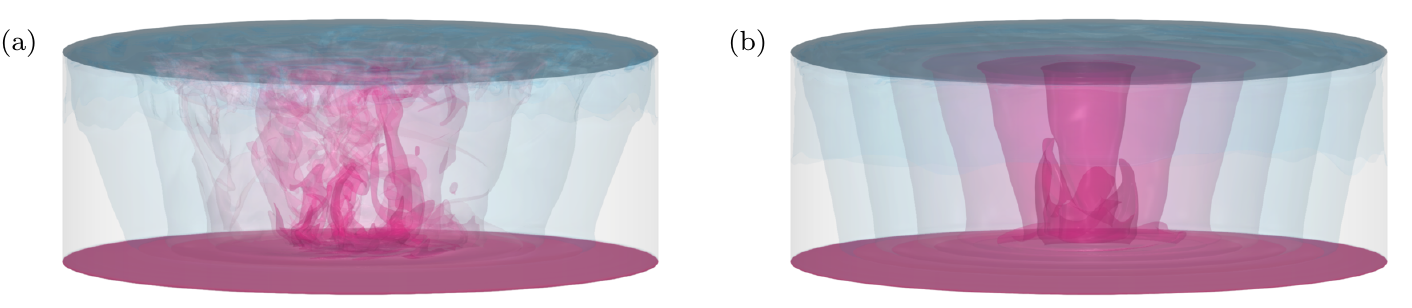}}
\caption{Temperature isosurfaces for $Ra = 10^8$, $Pr = 6.52$, $Ro_\parallel = 1.0$, and a larger aspect ratio of $\gamma = 1.502$. (a) $Fr= 1.5$ corresponding to a superfroudeality of $\chi = Fr/\gamma = 1$; (b) $Fr = 4.1$ corresponding to $\chi = 3$.  \label{fig:large}}
\end{figure*}

Figure~\ref{fig:large} shows two cases for $Ro_\parallel = 1.0$ and a four times larger aspect ratio of $\gamma = 1.5$. These solutions provide visual evidence that multiple vortices become more likely and the sheaths feature becomes more pronounced for larger $\gamma$. The higher aspect ratio shifts the transition to centrifugally dominated dynamics according to eq.~\eqref{eq:gamma}. Thus, the vortices in figure~\ref{fig:large} (a) at $Fr = 1.5$, i.e. at the transition border, are less tornado-like than the ones for $\gamma = 0.365$ at much lower $Fr$ and the same $Ro_\parallel$. However, they are orbiting around a common centre, again similar to outbreaks where multiple tornadoes are spawned within the very same mesocyclone. For $Fr = 4.1$, shown in figure~\ref{fig:large} (b), a single strong central vortex develops with tornado-like surrounding sheaths.

\subsection{Tornado-Like Vortices at $Ro_\parallel = 1.0$ and $Fr = 1.0$}
 \begin{figure}[t!]
\includegraphics[width=\textwidth]{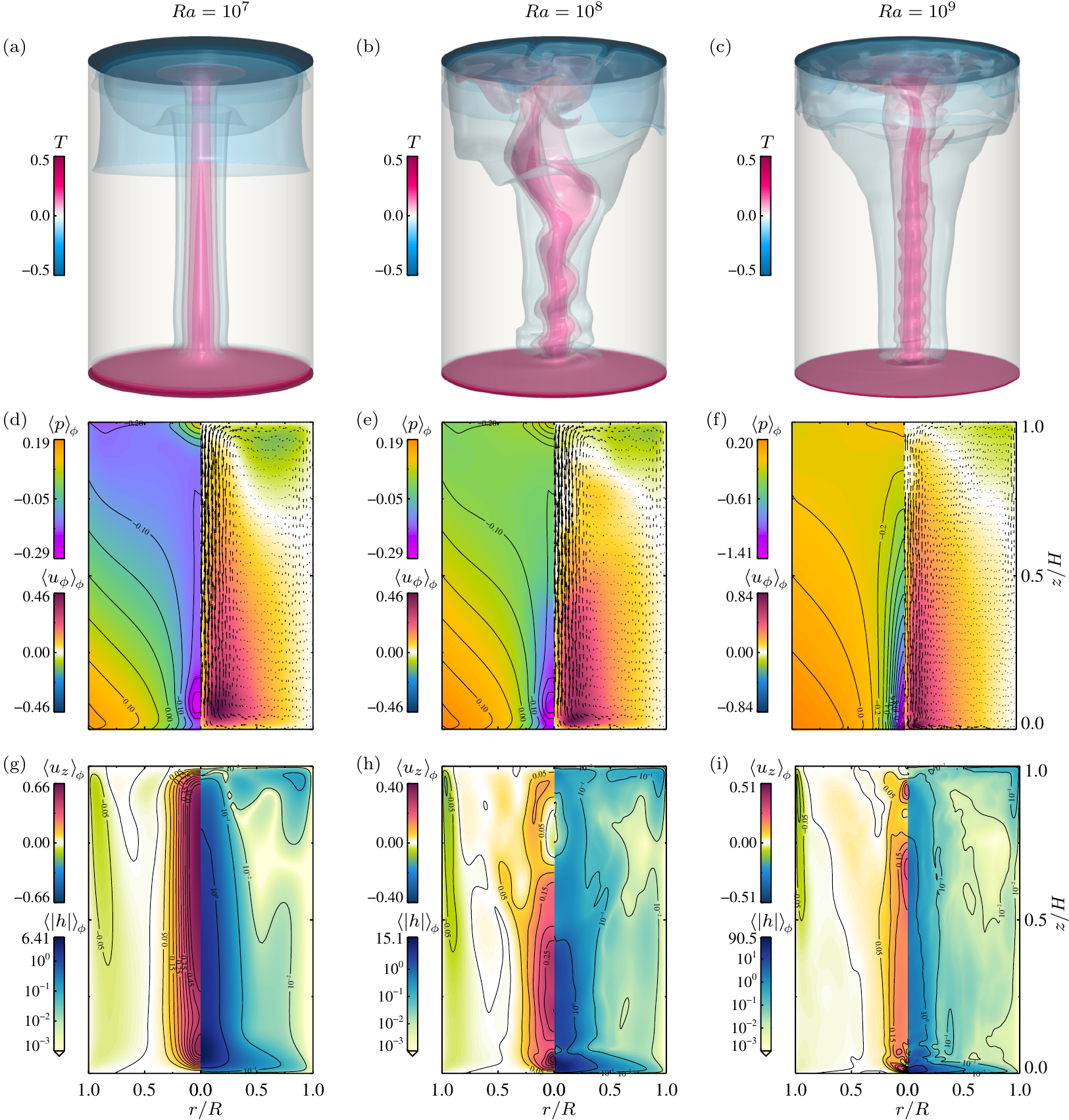}
\caption{Instantaneous flow fields for $Ro_\parallel = 1.0$, $Fr = 1.0$, $Pr = 6.52$, $\gamma = 0.365$, which yields the dependent alternative parameters $Ro_\perp = 1.654$, $S = 0.5$ and $\chi = 2.74$. The left column (a, d, g) shows data for $Ra = 10^7$, and, consequently, $Ek = 8.08 \times 10^{-4}$, $Re = 1.24 \times 10^3$, and $Re_v =  619$. The middle column (b, e, h) shows data for $Ra = 10^8$, and, hence, $Ek =  2.55 \times 10^{-4}$, $Re = 3.92 \times 10^3$, and $Re_v =  1958$. The right column (c, f, i) shows data for $Ra = 10^9$, and, hence, $Ek = 8.08 \times 10^{-5}$, $Re = 1.24 \times 10^4$, and $Re_v = 6192$.
In the top row (a--c) the three-dimensional temperature field is visualized employing ten isocontours between $T_b = 0.5$ and $T_t = -0.5$. In the middle and bottom row (d--f) azimuthally averaged fields at the same instant in time are presented, the $\phi$-average assists in clarifying the behavior around the centre $r=0$, where the tornado-like vortex is formed. In the middle row (d--f), the left contour plots shows the full range of the reduced pressure $p$, which is normalized such that $\langle p(r=R, z=H/2) \rangle_\phi = 0$ \citep[cf.][]{Fiedler1994}. The right contour plots show the azimuthal velocity $u_\phi$ in the range $[-\max|u_\phi|,\max|u_\phi|]$, the radial and vertical velocity vectors are overplotted, showing the meridional circulation. The lower row (g--i) shows contours of the vertical velocity $u_z$ on the left side, in the range  $[-\max|u_z|,\max|u_z|]$, and the absolute helicity $|h| = |\vec{\omega} \cdot \vec{u}|$ on a logarithmic scale in the range $[10^{-3},\max|h|]$ on the right side. \label{fig:flow}}
 \end{figure}

The following analysis focuses on the $\gamma = 0.365$ DNS with the tornadic case at $Ro_\parallel = 1.0$ and $Fr = 1.0$ being discussed in greater detail. In figure~\ref{fig:flow}, the relevant flow variables for this particular case are shown for three different Rayleigh numbers $Ra = 10^7$, $Ra = 10^8$, and $Ra = 10^9$, equivalent to the three different Reynolds numbers $Re = 1.24 \times 10^3$, $3.92 \times 10^3$, and $Re = 1.24 \times 10^4$. 

The instantaneous three-dimensional temperature $T$ is shown in figure~\ref{fig:flow} (a--c). In addition, the pressure $p$, the azimuthal velocity $u_\phi$ with overplotted velocity vectors, the vertical velocity $u_z$, and the absolute helicity $|h| = |\vec{\omega} \cdot \vec{u}|$ are shown as azimuthal averages $\langle \cdot \rangle_{\phi}$ in figure~\ref{fig:flow} (d--i). Despite the two higher $Ra$ cases not being fully axisymmetric, the azimuthal average is chosen to provide a view into the interior behaviour and to ease the comparison of the vortex core and the surrounding and sidewall adjacent properties.

The three-dimensional temperature fields reveal prototypical TLVs for all considered $Ra$. There is a single warm, central, columnar vortex that posseses so-called bulging deformations \citep{Ward1972} as well as secondary instabilities in form of spiraling bands \citep{Wurman1996}. The number of both the bulges and the helical instabilities increases, whereas their size and wave length decrease with increasing $Ra$. In fact, the $Ra = 10^7$ case in figure~\ref{fig:flow} (a) is quasi-steady and only has one bulge and no helical secondary instabilities. The core size itself exhibits at most a mild dependence on $Ra$. Hence, in that respect our thermally driven vortices behave comparably to mechanically forced vortices in a tornado simulator \citep{Davies-Jones1973}.
The $Ra = 10^8$ and $10^9$ cases in figure~\ref{fig:flow} (b) and (c), respectively, also demonstrate clear signs of a vortex break down in the upper part of the domain, where the central vortex becomes broader and more turbulent, similar to a hydraulic jump. For $Ra = 10^7$ the Rayleigh number is too low and, thus, the vortex extends from the bottom up to the top where it impinges and smoothly feeds into the upper recirculation. 
In line with this, the vertical velocity $u_z$ within the TLV is less intense, and even negative $u_z$ for $Ra = 10^8$ and $10^9$, as shown in figure~\ref{fig:flow} (d--i). Such downdrafts and the hereby induced formation of two-celled vortices are well-known features in tornadoes as well as numerically and laboratory produced TLVs \citep{Fiedler1986, Fiedler1994, Nolan2012, Rotunno2016, Trapp1999}.

The meridional overturning circulation induced by centrifugal buoyancy \citep{Curbelo2014, Horn2018, Hart1999, Hart2000, Marques2007} is not only the source of the warm central upwelling tornado-like vortex but, due to continuity, also of the cold downwelling along the sidewalls. Thus, the fluid immediately surrounding the vortex descends, analogous to what is observed in nature \citep{Marquis2012}. The cold downflow can wrap around the TLV as readily seen for $Ro_\parallel = 0.5$ and $0.5 \leq Fr \leq 2.0$ in figure~\ref{fig:tornadoes}. This wrapping effect will appear even stronger when viewed from the stationary external frame, and would also be visible for the $Ro_\parallel = 1.0$ cases. This qualitatively agrees with observations during the tornadic phase in a supercell thunderstorm, although the downdraft is additionally accompanied by heavy precipitation, effects that have been neglected here \citep{Markowski2014, Brandes1984, Dowell2002a}.

The radial temperature gradients that develop in our DNS are associated with a pressure drop in the central region (left panels of figure~\ref{fig:flow} (d--f)). More precisely, a ring--like quasi-axisymmetric pressure pattern is observed close to the bottom boundary.
The azimuthally averaged pressure is depicted in figure~\ref{fig:flow} (d--f), left panel, where it is normalized such that at $\langle p \rangle_\phi$ equals zero at the sidewall $r = R$ and half-height $z = H/2$. In the right panel the corresponding velocity fields is visualized. The contours show $\langle u_\phi \rangle_\phi$ and the vector arrows indicate the radial and vertical velocity components $(\langle u_r \rangle_\phi, \langle u_z \rangle_\phi)$. 
Since our governing eqs.~\eqref{eq:NSnd1}--\eqref{eq:NSnd3} describe the co-rotating frame of reference, the imposed rotation rate needs to be taken into account when comparing to velocities measured, e.g., with mobile Doppler systems. 
For clarity, the axial velocity $\langle u_z \rangle_\phi$ is also shown separately in the left panels of figure~\ref{fig:flow} (g--i).
Both the axial and the azimuthal velocity have comparable high magnitudes. Thus, centrifugal buoyancy consistently reconciles the high wind speeds with low pressure, which is often an issue with tornado models \citep{Fiedler1994, Fiedler1995}. The azimuthal wind speeds as well as the pressure drop show an increase with $Ra$, whereas the axial velocity remains approximately constant. 

Thus, the self-consistently generated TLVs in the QC regime of C$^{3}$ possess all of the three tornado-defining features outlined by Ward \cite{Ward1972}, and more. Other common tornado characteristics are reproduced as well. For all Rayleigh numbers, we observe a ``drowned vortex jump'' close to the bottom boundary layer, a well-known feature of tornadoes including dust-devils \citep{Maxworthy1973}. This means the radial boundary layer flow coming from the high pressure outer region overshoots into the central low-pressure region and then tilts upwards. This also creates an eye-like structure enclosed in the flow which is visible for all $Ra$ in figure~\ref{fig:flow}. Eyes have also been observed in the majority of strong tornadoes with Doppler radar observations, and unlike in hurricanes, the eyes do not necessarily extend through the entire storm \citep{Wurman1996, Lewellen1993, Oruba2017}. Our DNS reveal that the eye becomes stronger the closer the flow is to the CC regime \citep{Horn2018}. This implies that pronounced eye structures preferentially exist in a quasi-gradient balance, when pressure gradient, Coriolis, and centrifugal forces are in a triple balance \citep{Horn2018, Willoughby1990}.  

\subsection{Azimuthal Velocity Profiles and Maximum Wind Speeds}
\begin{figure}[t!]
\includegraphics[width=\textwidth]{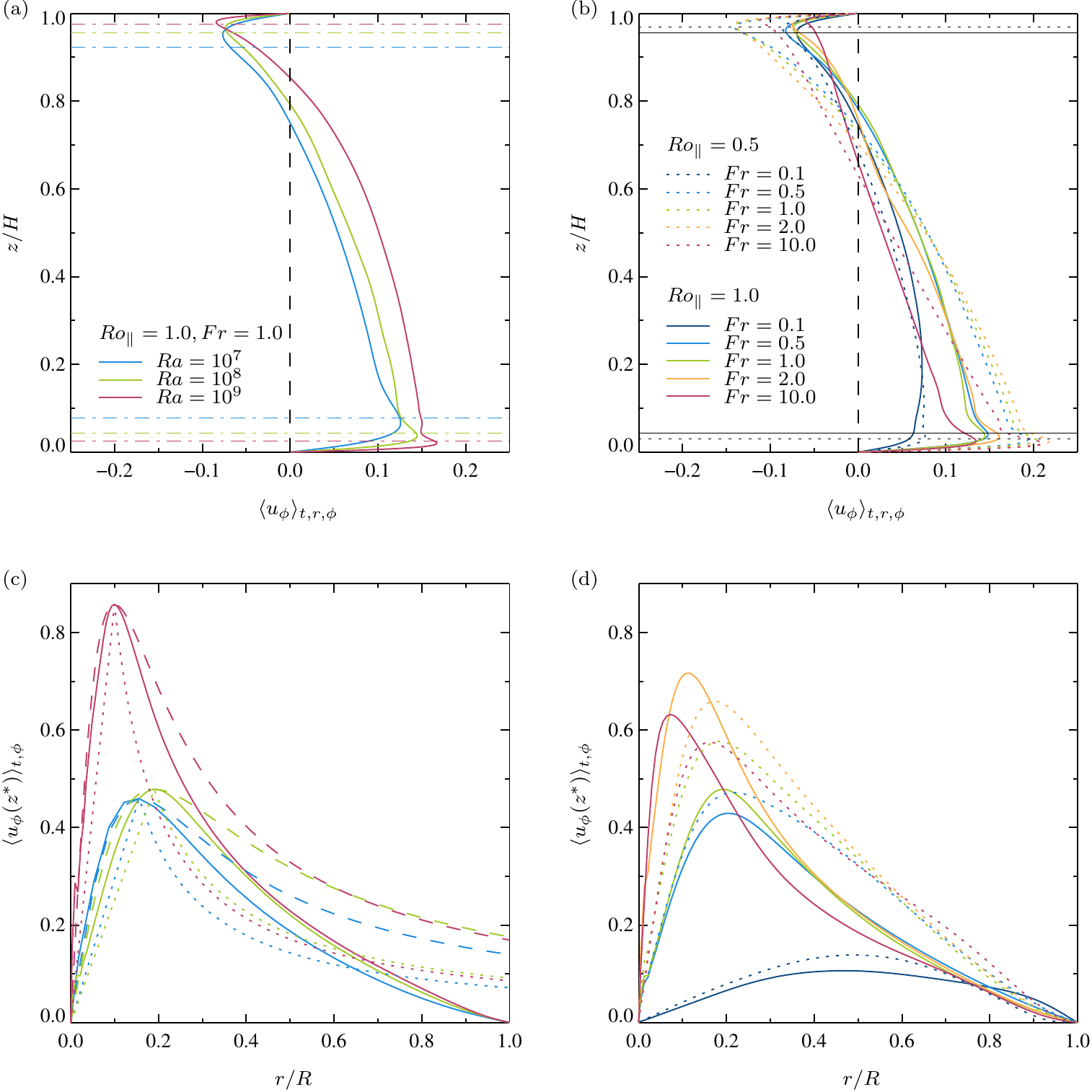}
\caption{Azimuthal velocity profiles for the tornado-like solution. (a) Vertical profiles of the temporally, radially, and azimuthally averaged profiles $\langle u_\phi \rangle_{t,r,\phi}$ for constant $Ro_\parallel = 1.0$ and $Fr = 1.0$, and varying Rayleigh numbers, $Ra = 10^7$ (blue), $Ra = 10^8$ (green), and $Ra = 10^9$ (purple). The bottom Ekman boundary layers, $\lambda_{Ek} = Ek^{1/2}$ are demarcated by dash-dotted lines in the same color. (b) Similar as figure~(a) but for constant $Ra =10^8$ and $Fr$ varying between $0.1$ to $10.0$ for $Ro_\parallel = 0.5$ (dotted lines) and $Ro_\parallel = 1.0$ (solid lines), respectively.
(c) Radial profiles of the temporally and azimuthally averaged profiles  $\langle u_\phi(z^*) \rangle_{t,r,\phi}$ evaluated at the $z$-value of the maximum of the profiles in figure~(a). In addition the dotted line shows the theoretical profile of a Rankine vortex, eq.~\eqref{eq:rankine}, and the dashed line the profile of a Fiedler vortex, eq.~\eqref{eq:fiedler}. (d) Corresponding radial profiles of the temporally and azimuthally averaged profiles  $\langle u_\phi(z^*) \rangle_{t,r,\phi}$ evaluated at the $z$-value of the maximum of the profiles in Fig. (c). \label{fig:profiles}} 
\end{figure}

In natural tornadoes, the highest wind speeds occur evidently in the azimuthal velocity component $u_\phi$, and time-averaged profiles are frequently sought with Doppler radar scans \cite{Tanamachi2007, Wurman2012, Bluestein1993}. Thus, we have evaluated the $u_\phi$ profiles of our C$^3$ DNS, shown in figure~\ref{fig:profiles}. These profiles can also be used to quantify the size of the eye and with it the position of the maximum wind speed. 

In figure~\ref{fig:profiles}~(a), the profiles $\langle u_\phi \rangle_{t, r, \phi}$ are averaged in time, radial and azimuthal direction and plotted as a function of the vertical coordinate $z$, for fixed $Ro_\parallel = 1.0$ and $Fr = 1.0$ and $Ra = 10^7$, $10^8$, and $10^9$. Unlike in $Fr=0$ cases, the profiles show a strong top-bottom asymmetry. 
This asymmetry expresses itself in a zero-crossing of the profiles occurring closer to the top boundary at $z/H \approx 0.8$ instead of at $z/H = 0.5$. Further, the absolute value of the maximum $\langle u_\phi \rangle_{t, r, \phi}$ value is about twice as high as that of the corresponding minimum. 
The maximum corresponds to the point where the TLV is swirling most rapidly and accordingly where the tornado is most destructive. The upper minimum corresponds to the weaker retrograde outer circulation, also visible in figure~\ref{fig:flow} (d--f).

The maximum of the velocity profiles occurs at $z^* \approx Ek^{1/2}$, i.e. at the edge of a classically derived Ekman boundary layer thickness $\lambda_{Ek}$ \citep{Greenspan1968}. The positions of the minimum near the top boundary exhibit, however, slight deviations from the Ekman layer behavior at lower $Ra$. Figure~\ref{fig:profiles}~(b) shows the vertical profiles for fixed $Ra = 10^8$, the gravitational Rossby numbers $Ro_\parallel = 0.5$ and $1.0$, respectively, and for varying Froude numbers, $0.1 \leq Fr \leq 10.0$. The maxima of the  $\langle u_\phi\rangle_{t,r,\phi}$ profiles are again located at $z^* \approx Ek^{1/2}$, thus, suggesting that there is no significant Froude dependence of $z^*$ in the tornado-relevant parameter range. 

We also evaluated the radial profiles averaged in time and in azimuthal direction at the height of the maximum wind speed, $\langle u_\phi(z^*) \rangle_{t,\phi}$. They are depicted in figure~\ref{fig:profiles}~(c) for fixed $Ro_\parallel = 1.0$ and $Fr = 1.0$ and the three Rayleigh numbers $Ra = 10^7$, $10^8$, and $10^9$. All three profiles exhibit the same qualitative behavior, but the maximum $u_\phi$ for $Ra = 10^9$ is almost twice as high as for $10^7$ and $10^8$. In the core region, $r \leq r^*$, i.e. at radial positions smaller than the radius $r^*$ of the wind maximum $u_\phi^*$, the profiles follow a quasi-linear relationship that has also been observed by Doppler measurements in real tornadoes \citep{Wurman2012}. In the outer region, $r > r^*$, the profiles exhibit a shallower decay. Thus, the wind speed profiles compare favorably with the standard tornado models, the Rankine vortex and the Fiedler vortex \citep{Fiedler1989, Fiedler1994, Dowell2005}, also shown in figure~\ref{fig:profiles}~(c). The Rankine vortex (RV) is defined as
\begin{equation}
u_\phi^{RV} = \left\{ \begin{array}{ll} \displaystyle \frac{u_\phi^* \, r}{r^*}& r \leq r^*\\[1em] \displaystyle \frac{u_\phi^* \, r^*}{r} & r > r^* \end{array} \right.,  \label{eq:rankine}
\end{equation}
i.e. a solid body rotation inside the vortex core, $r \leq r^*$, up to the radius of maximum wind speed $u_\phi^*$ and a potential flow outside for $r > r^*$.  The Fiedler vortex (FV) is an advanced version of the RV that is smoothly continuous at $r^*$ and is defined as 
\begin{equation}
u_\phi^{FV} = \frac{2 \, u_\phi^* r^* r}{r^{*2} + r^2}. \label{eq:fiedler} 
\end{equation}
The difference in the decay behavior between our DNS profiles and the RV and FV are due to the no-slip condition we employ at the sidewall that enforces $u_\phi = 0$ at $r = R$, as well as the relatively small aspect ratio $\gamma$.

Figure~\ref{fig:profiles}~(d) shows the radial profiles $\langle u_\phi(z^*) \rangle_{t,\phi}$ corresponding to the vertical profiles of figure~\ref{fig:profiles}~(b) for fixed $Ra = 10^8$ and $Ro_\parallel = 0.5$ and $1.0$, respectively, and varying $0.1 \leq Fr \leq 10.0$. Characteristic tornado profiles only develop for $Fr \gtrsim \gamma$. However, the maximum windspeed depends on both  $Ro_\parallel$ and $Fr$, with the highest $u_\phi^* = 0.72$ occurring for $Ro_\parallel = 1.0$ and $Fr = 2.0$.

\begin{figure}
\includegraphics[width=\textwidth]{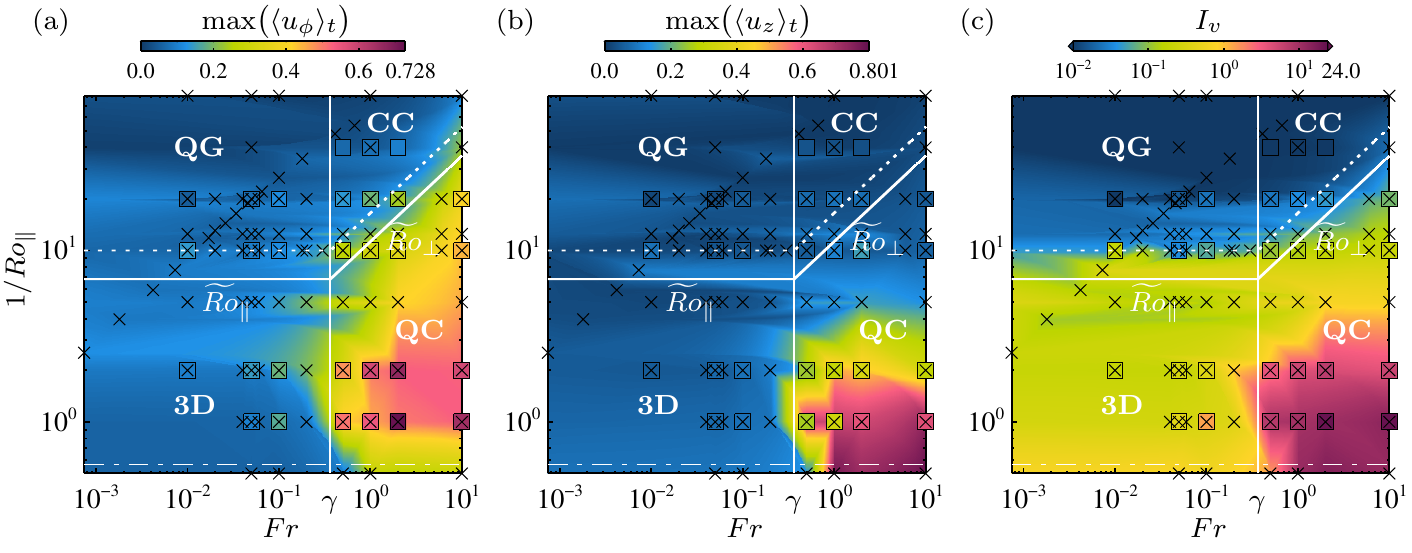}
\caption{(a) Maximum value of the temporally averaged azimuthal velocity $\max(\langle u_\phi \rangle_{t})$ in the full simulated $Fr- Ro_\parallel$ space based on the data for $Ra = 10^7$ (black crosses). In addition, the color-filled symbols show the data for $Ra =10^8$ using an identical color-code. The phase space is divided into the three-dimensional (3D), quasi-geostrophic (QG), quasi-cyclostrophic (QC), and the Coriolis-centrifugal (CC) regime. The boundaries between the 3D and QG regime are given by eq.~\eqref{eq:Ropar} and marked with the horizontal solid and dotted white lines for $Ra = 10^7$ and $Ra = 10^8$, respectively. The boundaries between the QC and the CC regime according to eq.~\eqref{eq:Roperp} are marked with the slanted corresponding lines. The white vertical solid line indicates the line $Fr = \gamma$ that separates the non-centrifugal regimes (3D and QG) and the centrifugally dominated regimes (QC and CC). None of these transitions are expected to be sharp \citep[cf.][]{Horn2018, Horn2019}. In addition, the thin horizontal dash-dotted line indicates the bifurcation \eqref{eq:Weiss} to mildly Coriolis affected flows \citep{Weiss2010}. The tornado-like solutions are situated in the QC regime. (b) Maximum value of the temporally averaged vertical velocity $\max(\langle u_\phi \rangle_t)$. (c) Intensification of the vortex strength $I_v$, defined as ratio of the maximum azimuthal velocity $U_\phi^*$ to the radius where it occurs $R^*$,  normalized by the ambient rotation $\Omega$; in the here used non-dimensionalization, this means $I_v \equiv 2 U_\phi^* Ro_\parallel / (R^* \sqrt{\gamma})$. \label{fig:contours}}
\end{figure}

The full parameter space is given in figure~\ref{fig:contours} in form of a contour map to further assess the dependence of the maximum azimuthal and vertical windspeed, $\max(\langle u_\phi \rangle_t)$ and $\max(\langle u_z\rangle_t)$, respectively, on $Ro_\parallel$, $Fr$ and $Ra$. Figure~\ref{fig:contours}(a) shows that the strongest enhancement of $\max( \langle u_\phi \rangle_t)$ occurs in the QC regime, bounded by our predictions for $\widetilde{Ro}_\perp$ and $\widetilde{Fr}$, as well as $\widetilde{Ro}{}_\parallel^W$  as discussed in  \textsection~\ref{sec:c3}. In the 3D, QG, and CC regime, $\max( \langle u_\phi \rangle_t)$ is insignificant. This supports our hypothesis that a quasi-cyclostrophic balance is neccessary for TLVs to form.

Instantaneously, comparably high values of $u_\phi$ can also be obtained in the 3D regime and at the edge towards the QG regime. 
These high azimuthal velocities are associated with detaching plumes and swirling Ekman vortices \citep{Stevens2009, Weiss2010}, where, $u_\phi$ is not spatially localized and also not stable over long time periods. The maximum vertical windspeed $\max(\langle u_z\rangle_t)$, shown in figure~\ref{fig:contours}(b), is determined by centrifugal buoyancy effects alone. Thus, $Fr$ is required to be greater than $\gamma$, and the enhancement of $u_z$ also occurs for $Ro_\parallel = \infty$, as was already visually suggested by figure~\ref{fig:tornadoes}. The minimum $Ro_\parallel$ for which the $u_z$ enhancement is observed, however, appears to be rather set by $\widetilde{Ro}_\parallel$ and not by $\widetilde{Ro}_\perp$, and a similar disparity was also found in the behavior of the plane averaged and point wise centre temperatures \citep{Horn2019}. The underlying reason is still unclear.

\subsection{Vortex Intensification and Helicity Considerations}
To connect these results further to actual tornadoes we also consider the vortex intensification $I_v$ defined as the ratio of the maximum temporally averaged azimuthal velocity 
\begin{equation}
 U_\phi^* \equiv \max \bigl( \langle u_\phi \rangle_t \bigr)
\end{equation}
to the radius where it occurs $R^*$ and the ambient rotation $\Omega$. With our non-dimensionalisation, this yields
\begin{equation}
I_v \equiv \frac{2 U_\phi^* Ro_\parallel}{R^* \sqrt{\gamma}},
\end{equation}
and the phase diagram is shown in figure~\ref{fig:contours}(c). $I_v$ may also be understood as a local swirl ratio, or a local vortex-scale Rossby number. In nature, this translates to how much a tornado's vorticity is intensified relative to the ambient mesocyclonic vorticity. The maximum intensification occurs within the QC regime.
In contrast, in the 3D regime, $I_v$ is negligible, and in the QG and CC regimes, a diminishment of an order of magnitude and lower occurs. In the DNS, the highest intensification is $24.0$ and, thus, between one to two orders of magnitude higher than the applied $\Omega$. This agrees with the values obtained in natural tornadic settings \citep{Lin1992, Snow1987}. The larger value of $U_\phi^*$ obtained for $Ra = 10^9$ (cf. figure~\ref{fig:profiles}) is suggestive that for higher $Ra$, in particular atmospheric $Ra$ values, stronger intensifications $I_v$ can occur.  

Similar to real tornadoes and supercell thunderstorms \citep{Klemp1987, Lilly1986b}, we also find that the TLVs in the QC regime of C$^3$ optimise the helicity $h = \vec{\omega} \cdot \vec{u}$. For all $Ra$, visualized in the right panels of figure~\ref{fig:flow} (g--i), $h$ is highest close to the bottom where the vortices detach and along the vortex axis, but also the recirculation at the top outer rim is associated with relatively high magnitudes in $h$. The helicity increases with $Ra$, in particular, due to the developing secondary helical instabilities on the vortex core. The maximum values of $h$ are about twice and fourteen times higher for $Ra = 10^8$ and $Ra = 10^9$, respectively, than for $Ra = 10^7$. 

\begin{figure}
\centerline{\includegraphics[width=\textwidth]{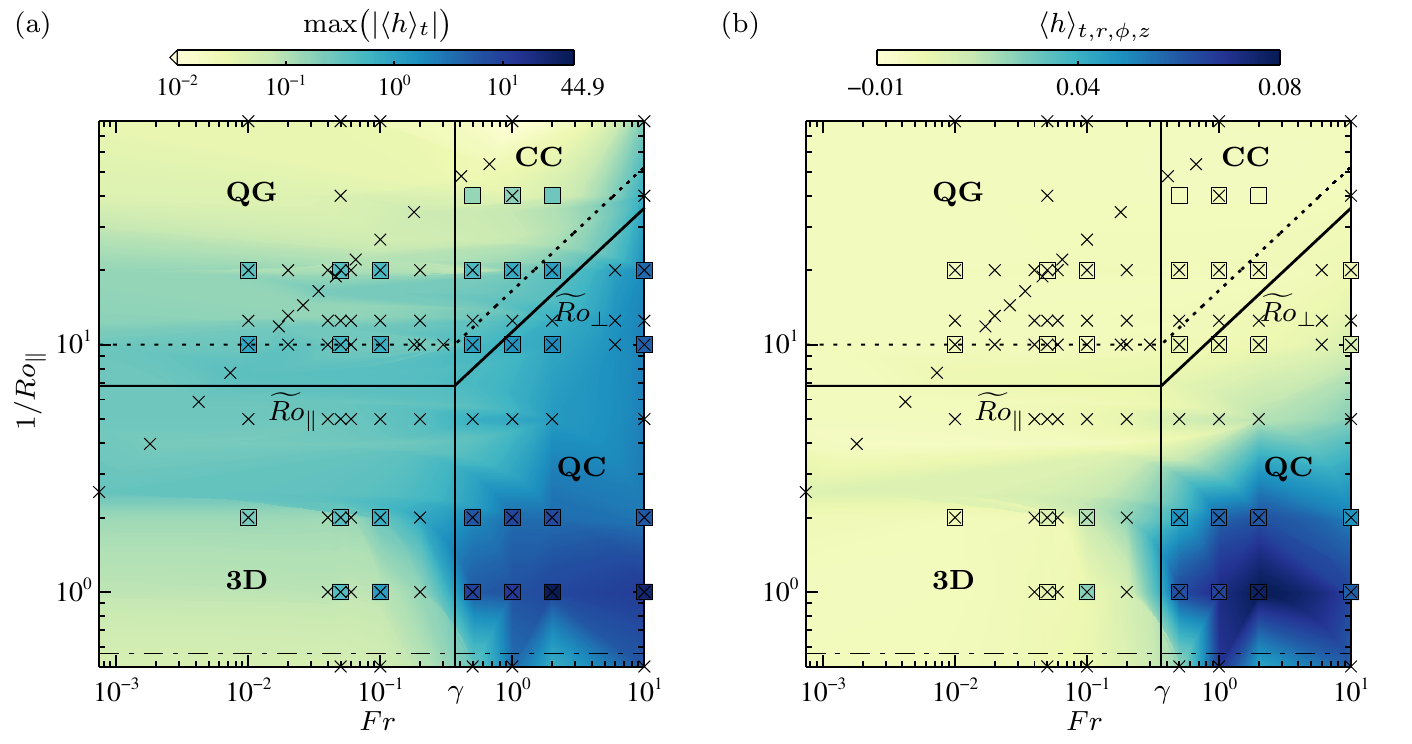}}
\caption{(a) Maximum value of the temporally averaged absolute helicity $|\langle h \rangle_t| =|\langle \vec{\omega} \cdot \vec{u}\rangle_t|$ on a logarithmic scale. (b) Volume averaged helicity $\langle h \rangle_{t,r,\phi,z}$. See figure~\ref{fig:contours} for a description of the regime borders.  \label{fig:helicity}}
\end{figure}

Figure~\ref{fig:helicity} (a) displays the phase diagram of the absolute helicity $|\langle h \rangle_t| =|\langle \vec{\omega} \cdot \vec{u}\rangle_t|$ and figure~\ref{fig:helicity} (b) the phase diagram of the volume averaged helicity $\langle h \rangle_{t,r,\phi,z}$. 
Both,  $|\langle h \rangle_t|$ and  $\langle h \rangle_{t,r,\phi,z}$, are significantly higher in the tornadic QC regime, whereas they are negligible in the 3D, QG, and CC regime. Since a high helicity reduces dissipation by delaying the turbulent energy cascade \citep{Lewellen1993}, high helicity vortices decay much more slowly than low helicity vortices.  
This further explains the very coherent TLVs that develop in the QC regime.
  
\subsection{Influence of the Mechanical Bottom Boundary Conditions\label{sec:bcs}}
Despite the many commonalities of the TLVs obtained in our DNS and natural tornadoes, C$^3$ is still a tremendously simplified model system. The most significant discrepancies are likely the mechanical boundary conditions, especially at the bottom boundary. In our model, we have a closed container with no-slip walls that rotate at angular velocity $\Omega$. On the other hand, a natural tornado is formed within a rotating mesocyclone that is bounded at the bottom by the non-moving stationary ground. We will not address the issue of the lateral boundary conditions, but only focus on the bottom boundary condition. 

\begin{figure}[t!]
 \includegraphics[width=\textwidth]{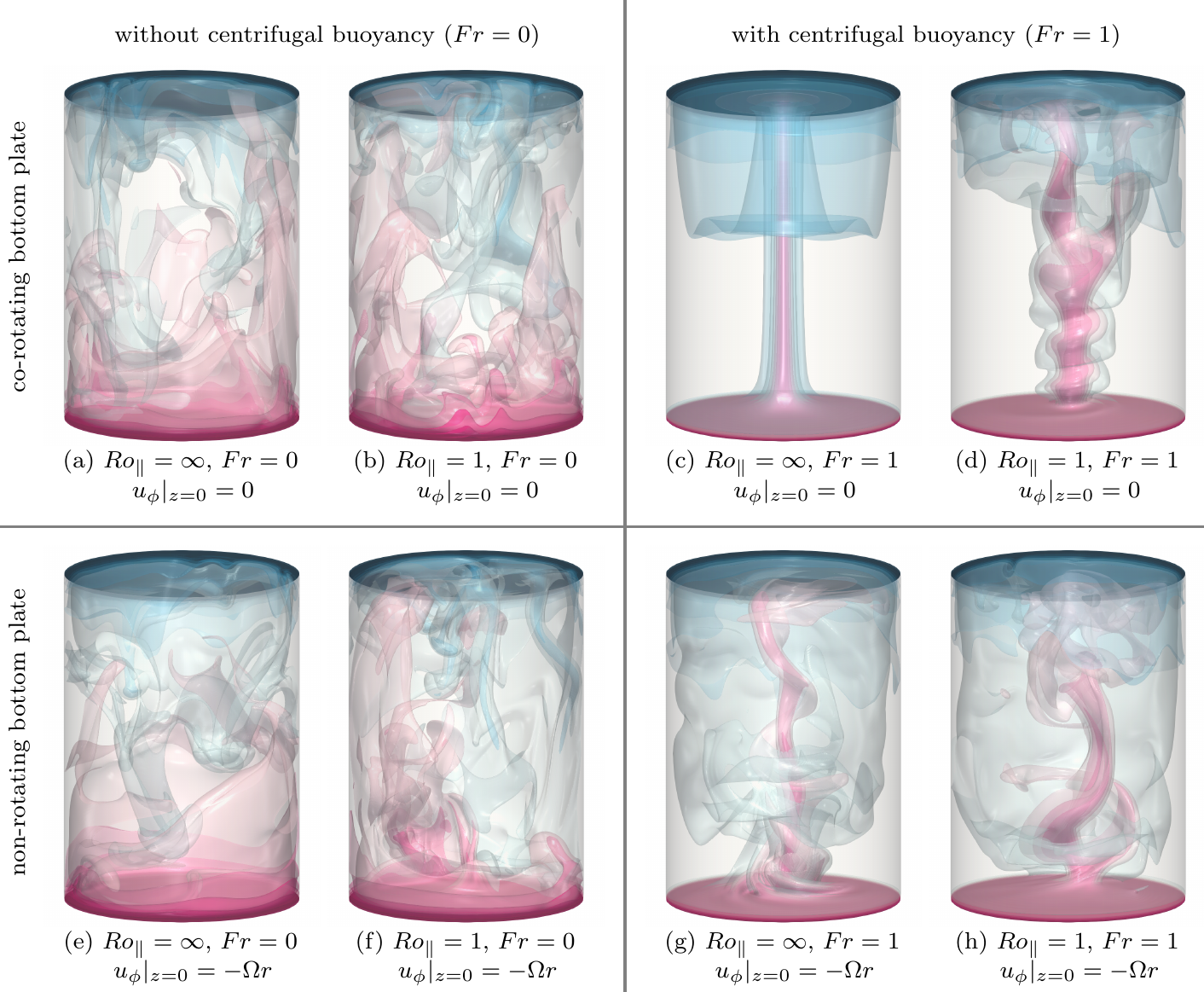}
 \caption{Modeling suite at $Ra = 10^8$ that tests the effects of Coriolis forces, centrifugal bouyancy forces, and differing bottom plate mechanical boundary conditions on the formation of TLVs. The panels on the left (a, b, e, f) correspond to cases without centrifugal buoyancy ($Fr=0$) and the ones on the right (c, d, g, h) correspond to cases that include centrifugal buoyancy ($Fr = 1$). The simulations shown in the upper panels (a--d) are with the default co-rotating bottom boundary conditions. The lower panels (e--h) have a non-rotating bottom plate to mimic a more realistic stationary ground. These simulation results show that centrifugal buoyancy is the essential component for TLV-genesis in the present C$^3$ system.\label{fig:bcs}}
\end{figure}

Since the DNS here are in the co-rotating frame of reference, we prescribe an oppositely rotating angular velocity at the lower boundary, i.e. $u_\phi(r,\phi, z = 0) = - \Omega r$ to simulate the non-rotating stationary ground. We choose four different representative cases at $Ra = 10^8$, namely, $(Ro_\parallel, Fr) = \{(\infty, 0), \, (1,0), \, (\infty, 1), \, (1,1)\}$, and compare the co-rotating and non-rotating bottom plate boundary conditions, as shown in figure~\ref{fig:bcs}. The cases with $Fr = 0$, figure~\ref{fig:bcs} (a,b,e,f), where centrifugal buoyancy is absent, do not develop TLVs with either boundary condition. The cases with $Fr = 1$, figure~\ref{fig:bcs} (c,d,g,h), where significant centrifugal buoyancy is present, exhibit TLVs with both boundary conditions. 
The counter-rotating bottom plate, mimicking the non-rotating ground, introduces more irregular motions for $Ro_\parallel = \infty$ in figure~\ref{fig:bcs} (g) as it brings back a Coriolis forcing. This additional Coriolis forcing appears to promote a two-celled structure which can also be found in tornadoes \cite{Sullivan1959}.  

Based on the figure~\ref{fig:bcs} results, we argue that centrifugal buoyancy is the key ingredient to produce TLVs in the C$^3$ system. 

\section{Estimating Formation Conditions for Natural Tornadoes}
An essential question remains as to whether centrifugal buoyancy is of significant strength in natural tornadoes. This question can be answered by considering the local system of the rotating updraft, or mesocyclone, within a supercell thunderstorm, as outlined in the schematic of figure~\ref{fig:sketch}(b), and by estimating the non-dimensional control parameters.

Assuming the storm has reached the stage in which the near-surface cyclone has moved underneath the mesocyclone that is aloft, then we can consider a wide vortex column with a typical radius of $R_{mc} \simeq \SI{1.5}{km}$ to \SI{4.5}{km}. This column extends from the ground up to the height of the tropopause at about \SI{12}{km} \citep{Davies-Jones2015, Snow1987}, as shown in the figure~\ref{fig:sketch} (b) schematic. This gives an estimate of $\gamma_{mc} \simeq 0.125$ to $\sim 0.375$. According to eq.~\eqref{eq:gamma}, centrifugal buoyancy effects become significant for $Fr \lesssim 1$, since the aspect ratio $\gamma$ is of order one and smaller. Crucially, the rotation rate and therefore Froude number is not set by Earth's rotation rate, but is instead set by the rotation rate of the mesocyclone $\Omega_{mc}$. 
Using observationally approximated circulation strengths \citep{Davies-Jones2015} of $\Gamma_{mc} \sim 2\pi R_{mc}^2 \Omega_{mc} \simeq \SI{2 \times 10^5}{m^2 s^{-1}}$ to \SI{6 \times 10^5}{m^2 s^{-1}}  and $g = \SI{9.81}{m s^{-1}}$ yield an ambient rotation rate of the mesocyclone in the range $\Omega_{mc} $ between $1.6 \times \SI{10^{-3}}{s^{-1}}$ and $4.7 \times \SI{10^{-2}}{s^{-1}}$.  With these values, we estimate that $Fr_{mc} = \Omega_{mc}^2 R_{mc}/g \lesssim 0.83$ in natural mesocyclones. 
Centrifugal buoyancy effects should then come into play in mesocyclone convection and tornadogenesis dynamics since we estimate that $Fr_{mc} \gtrsim \gamma_{mc}$. Alternatively stated, natural mesocyclones can reach the superfroudeal regime, $\chi_{mc} \gtrsim 1$, in which C$^3$ simulations show that centrifugal buoyancy controls TLV formation.

Additionally, the temperature is one of the crucial parameters determining tornadogenesis \citep{Markowski2014}. Because most tornadoes form in late spring and summer and at the time of day when the surface temperatures are highest, we assume ground surface temperatures between \SI{20}{\degC} and \SI{30}{\degC}. Further, supercell thunderstorms are also accompanied by hail, thus higher elevation temperatures must be well below \SI{0}{\degC} \citep{Bluestein2013}, and cloud temperatures below \SI{-60}{\degC} have been measured during tornado outbreaks \citep{Fujita1985}. Thus, we argue that a sensible estimate for the vertical temperature difference is $\Delta_{mc} \sim \SI{90}{K}$. However, since the potential temperature is what controls the atmospheric dynamics \citep{Cushman2011, Spiegel1971}, we take $\Delta_{mc} \sim \SI{30}{K}$ by assuming a \SI{5}{K/km} moist lapse rate. Using the following constant material properties for air at \SI{15}{\degC} and \SI{1}{atm}, $\nu = 1.45 \times \SI{10^{-5}}{m^2 s^{-1}}$, $\alpha = 3.48 \times \SI{10^{-3}}{K^{-1}}$, $\kappa = 2.02 \times \SI{10^{-5}}{m^2 s^{-1}}$ \citep{Batchelor1967}, we estimate $Ro_{\parallel,mc} \gtrsim  0.11$ and $Ro_{\perp,mc} \gtrsim 0.16$, which together with $Fr_{mc} \gtrsim \gamma_{mc}$, are values characteristic of the QC regime. We note that this a purely thermal estimate.
Other buoyancy sources are not presently considered in our idealized C{$^3$} system, such as evaporation and condensation, hydrometeors, dust, etc. \citep{Doswell2004, Davies-Jones2003}.
Hence, the effective $\Delta_{mc}$ and also $Ro_{\parallel,mc}$ and $Ro_{\perp,mc}$ may all be higher, resulting in values for tornadoes that lie still deeper within the tornadic QC regime.

The estimated Rayleigh number is $Ra_{mc} \sim 6 \times 10^{21}$, making $Ra$ the control parameter with the greatest deviation between our DNS and natural tornadoes. We speculate that a higher $Ra$ will not fundamentally alter the results, but only lead to a higher level of turbulence and more pronounced secondary instabilities.

The transition predictions to the QC tornado-bearing regime enable us to identify mesocyclone characteristics that determine tornado formation, maintenance, and dissipation. The global quantities in the definition of the control parameters are assumed to represent the mesocyclone values: the mesocyclone's rotation rate $\Omega_{mc}$, its height $H_{mc}$ and its vertical (potential) temperature difference $\Delta_{mc}$. The predictions \eqref{eq:Ropar}--\eqref{eq:gamma} yield estimates for the critical angular rotation rate $\widetilde{\Omega}_{mc}$ and the critical temperature difference $\widetilde{\Delta}_{mc}$ at which tornadoes can exist. Since $Ro_\perp$ is given solely through $Fr$, $Ro_\parallel$, and $\gamma$, only two of the relations are needed for a unique solution.

The critical centrifugal Rossby number $\widetilde{Ro}_\perp$, eq.~\eqref{eq:Roperp}, can be recast in dimensional form as
\begin{equation}
\widetilde{\Delta}_{mc} \gtrsim 13 \frac{2^{3/2} \kappa}{\alpha  g^{1/4} \nu^{1/2}} \frac{1}{H_{mc}^{3/4}}.
\end{equation}
Inserting the material properties of air yields the critical temperature difference
\begin{equation}
\widetilde{\Delta}_{mc} \gtrsim \frac{\SI{31.7}{K~m^{3/4}}}{H_{mc}^{3/4}}. \label{eq:Deltamc}
\end{equation}
and with $H_{mc} = \SI{12}{km}$ we obtain $\widetilde{\Delta}_{mc}~\gtrsim~\SI{0.03}{K}$. 

The proposed transitions $\eqref{eq:Ropar}$ and $\eqref{eq:Roperp}$ are not yet possible to test numerically or experimentally for atmospheric values of $Ra$. The actual dependence of  $\widetilde{Ro}_\parallel$ and  $\widetilde{Ro}_\perp$ on $Ra$ might be weaker, which would lead to a higher $\widetilde{\Delta}_{mc}$.
If we conservatively extrapolate our current estimate of $\widetilde{Ro}_\perp \simeq 0.1$ as requirement for tornado existence in a real atmosphere, we obtain $\widetilde{\Delta}_{mc} \simeq 4/\alpha = \SI{11.5}{K}$. Hence, whilst a higher $\Delta_{mc}$ is advantageous for the formation of tornadoes, it does not appear to be the decisive factor.

The critical Froude number $\widetilde{Fr} = \gamma$, corresponding to a superfroudeality of $\chi = 1$, can be equivalently expressed in dimensional form as
\begin{equation}
\widetilde{\Omega}_{mc} \gtrsim \sqrt{\frac{g}{H_{mc}}}. \label{eq:Omegamc}
\end{equation}
Significantly, eq.~\eqref{eq:Omegamc} estimates the critical mesocyclonic angular rotation rate $\widetilde{\Omega}_{mc}$ relative to the mesocyclone height $H_{mc}$. 
Taking $H_{mc} = 12$ km, we 
obtain $\widetilde{\Omega}_{mc} \gtrsim 3 \times \SI{10^{-2}}{s^{-1}}$, which is within the range of estimated $\Omega_{mc}$ values \citep{Davies-Jones2015}. Eq.~\eqref{eq:Omegamc} is much more restrictive than eq.~\eqref{eq:Deltamc} and, thus, we argue crucial for the prediction of tornado occurrences.

Lastly, it should be noted that the non-dimensional control parameters are fixed in our DNS, whereas in nature, they vary with time.
We argue that when the atmospheric parameters change such that the mesocyclone system no longer resides within the QC regime, the tornado decays. This is in line with observations, e.g., that tornadoes decease once the temperature difference decreases \citep{Marquis2012}, and dissipate when they become separated from their parent mesocyclone \citep{Dowell2002b}. Additionally, the TLV formation in our DNS happened almost instantaneously along the entire axis and usually within $t \approx {\mathcal{O}}(\tau_{cb})$. Since we argue physically that centrifugal buoyancy initiates tornadogenesis, the relevant, radius-independent time scale is 
\begin{equation}
\tau_{cb} \simeq \sqrt { \frac{1}{\alpha \Delta_{mc} \Omega_{mc}^2 }} = \sqrt{\frac{1}{\alpha \Delta_{mc}} \frac{H_{mc}}{g}} \, . 
\end{equation}
For the parameter estimates given here, this implies formation times of order \SI{65}{s} in natural tornadoes. This $\tau_{cb}$ estimate is much shorter than formation time estimates based on the dynamic pipe effect and is consistent with recent estimates of tornadogenesis \citep{Bluestein2019, French2013, Houser2015, Houser2018, Trapp1999}. 

\section{Summary and Concluding Remarks}
Idealised tornado-like vortices (TLVs) have been successfully generated in laboratory and numerical settings. However, these models heavily rely on an additional mechanical forcing such as fans or prescribed buoyancy forces that do not exist in nature. We have shown that in the system of Coriolis-centrifugal convection (C$^3$) TLVs can be self-consistently generated without such artificial forces. Our DNS advocate the inclusion of centrifugal buoyancy $\vec{b}_\perp$ in the radial momentum equation as an important puzzle piece in unraveling tornado physics, with our results differing greatly if we only consider gravitational buoyancy $\vec{b}_\parallel$ in the vertical momentum equation \citep[cf.][]{Doswell2004}.

In our DNS a rotating cylinder represents the rotating updraft, or mesocyclone, within a supercell thunderstorm, analogous to other studies of tornadoes in isolation from their parent storm. Crucially, this means that the applied rotation rate in our DNS corresponds to the circulation of the mesocyclone $\bm \Omega_{mc}$ and not Earth's angular rotation rate. The temperature field is explicitly included in our simulations, unlike in most other idealised studies of tornadoes and TLVs, where the heated ground and the cool upper atmosphere are modeled by plane isothermal boundaries. The C$^3$ system naturally exhibits vertical as well as significant horizontal temperature gradients. Thus, gravitational buoyancy provides an updraft, whilst centrifugal buoyancy provides a radial inward force close to the bottom boundary and a radial outward force along the top boundary.

Within the Oberbeck-Boussinesq approximation it is possible to decouple the influence of the Coriolis force, whose strength can be expressed by the gravitational Rossby number $Ro_\parallel$ (or alternatively the Ekman number $Ek$) and the centrifugal buoyancy force, whose strength can be expressed by the Froude number $Fr$. Within the C$^3$ phase space spanned by $Ro_\parallel$ and $Fr$, tornado-like solutions reside within the weakly Coriolis-influenced, quasi-cyclostrophic (QC) regime, i.e. where the dynamics are dominated by centrifugal buoyancy, and the Coriolis force is secondary, but non-negligible. 

We argue that the idealized C$^3$ system can reproduce essential features present in natural tornadoes. These features include an intensification of the azimuthal wind speed of the central tornadic vortex by more than one order of magnitude, a strong updraft, an eye in the centre, and occasional central downdrafts. The TLVs in our DNS develop spiral bands wrapping around the vortex core, and descending fluid immediately surrounding it. We find characteristic pressure ring profiles with a strong drop towards the centre and associated with it a drowned vortex jump. For suitable parameter combinations we also observe sheaths orbiting the vortex base, as well as multiple tornadoes.

Importantly, the C$^3$ system is able to produce a similar richness of tornado morphologies as found in nature. Further, because of our no-slip conditions, surface friction and vorticity along the edge of the lower boundary layer are dynamically important. This is similar to nature \citep{Davies-Jones2015, Nolan1999}, but currently there the source of vorticity remains unknown \citep{Bluestein2019}. Based on our C$^3$ results, we postulate that the combined effects of strong gravitational and centrifugal buoyancies, together with weak Coriolis deflection and active Ekman layers are viable sources of tornadic vorticity production.

Our studies provide a possible explanation why seemingly similar mesocyclones may or may not spawn tornadoes. It is not just the absolute dimensional value of the vertical angular velocity $\Omega_{mc}$, but also the mesocyclone geometry, as well as the temperature difference $\Delta_{mc}$ that require consideration. Non-dimensionally, these mesocyclone characteristics are expressed by the Froude number $Fr$, the aspect ratio $\gamma$, and the gravitational Rossby number $Ro_\parallel$.
We find that tornadic solutions only exist within the quasi-cyclostrophic regime and its immediate border regions. The QC regime is bounded by a superfroudeality $\chi = Fr/\gamma \simeq 1$. This defines a critical mesocyclonic angular rotation rate above which tornadoes can form, $\widetilde{\Omega}_{mc} = \sqrt{g/H_{mc}}$ where $g$ is the gravitational acceleration and $H_{mc}$ is the height of the mesocyclone. Sufficiently high $Ro_\parallel$ and $Ro_\perp$ are also required for tornado formation, such that centrifugal buoyancy dominates gravitational buoyancy, which dominates Coriolis force. 

In future efforts, we will seek to test the validity of our tornadic QC regime in settings closer to that of a realistic atmosphere. Further, we will refine the accuracy of our predictions by including additional mesocyclone processes \citep[e.g.,][]{Orf2017, Rotunno2013, Davies-Jones2015}.

\section*{Acknowledgements}
The authors thank the two anonymous referees for constructive comments that improved the manuscript, and gratefully acknowledge the NSF-EAR Geophysics Program who supported this work via awards \#1547269 and \#1853196.  

\bibliographystyle{tfnlm}
\bibliography{froude}

\begin{thebibliography}{100}
\providecommand{\url}[1]{\normalfont{#1}}
\providecommand{\urlprefix}{Available from: }

\bibitem{Horn2018}
Horn~S, Aurnou~JM. {Regimes of Coriolis-Centrifugal Convection}. Phys Rev Lett.
  2018;\hspace{0pt}(120):204502.

\bibitem{Horn2019}
Horn~S, Aurnou~J. {Rotating convection with centrifugal buoyancy: Numerical
  predictions for laboratory experiments}. Phys Rev Fluids.
  2019;\hspace{0pt}4:073501.

\bibitem{Aurnou2015}
Aurnou~JM, Calkins~MA, Cheng~JS, et~al. {Rotating convective turbulence in
  Earth and planetary cores}. Phys Earth Planet Inter.
  2015;\hspace{0pt}246:52--71.

\bibitem{Aurnou2018}
Aurnou~JM, Bertin~V, Grannan~AM, et~al. Rotating thermal convection in liquid
  gallium: Multi-modal flow, absent steady columns. J Fluid Mech.
  2018;\hspace{0pt}846:846--876.

\bibitem{Chandrasekhar1961}
Chandrasekhar~S. Hydrodynamic and hydromagnetic stability. Oxford: Clarendon
  Press; 1961.

\bibitem{Cheng2018}
Cheng~JS, Aurnou~JM, Julien~K, et~al. A heuristic framework for next-generation
  models of geostrophic convective turbulence. Geophys \& Astrophys Fluid Dyn.
  2018;\hspace{0pt}112(4):277--300.

\bibitem{Ecke2014}
Ecke~RE, Niemela~JJ. {Heat transport in the geostrophic regime of rotating
  {Rayleigh}-B{\'e}nard convection}. Phys Rev Lett.
  2014;\hspace{0pt}113(11):114301.

\bibitem{Favier2014}
Favier~B, Silvers~L, Proctor~M. {Inverse cascade and symmetry breaking in
  rapidly rotating Boussinesq convection}. Phys Fluids.
  2014;\hspace{0pt}26(9):096605.

\bibitem{Featherstone2016}
Featherstone~NA, Hindman~BW. The emergence of solar supergranulation as a
  natural consequence of rotationally constrained interior convection.
  Astrophys J Lett. 2016;\hspace{0pt}830(1):L15.

\bibitem{Horn2014a}
Horn~S, Shishkina~O. {Rotating non-Oberbeck--Boussinesq {Rayleigh}--B{\'e}nard
  convection in water}. Phys Fluids. 2014;\hspace{0pt}26(5):055111.

\bibitem{Horn2015}
Horn~S, Shishkina~O. {Toroidal and poloidal energy in rotating
  {Rayleigh}--B\'enard convection}. J Fluid Mech.
  2015;\hspace{0pt}762:232--255.

\bibitem{Horn2017}
Horn~S, Schmid~PJ. {Prograde, retrograde, and oscillatory modes in rotating
  Rayleigh--B\'enard convection}. J Fluid Mech. 2017;\hspace{0pt}831:182--211.

\bibitem{Julien1996c}
Julien~K, Legg~S, McWilliams~J, et~al. {Rapidly rotating turbulent
  {Rayleigh}--B{\'e}nard convection}. J Fluid Mech.
  1996;\hspace{0pt}322:243--273.

\bibitem{Julien2012}
Julien~K, Knobloch~E, Rubio~AM, et~al. {Heat transport in low-Rossby-number
  {Rayleigh}-B{\'e}nard convection}. Phys Rev Lett.
  2012;\hspace{0pt}109(25):254503.

\bibitem{Kunnen2008b}
Kunnen~RPJ, Clercx~HJH, Geurts~BJ. Enhanced vertical inhomogeneity in turbulent
  rotating convection. Phys Rev Lett. 2008 Oct;\hspace{0pt}101:174501.

\bibitem{Kunnen2013}
Kunnen~RPJ, Clercx~HJH, van Heijst~GF. {The structure of sidewall boundary
  layers in confined rotating {Rayleigh}--B{\'e}nard convection}. J Fluid Mech.
  2013;\hspace{0pt}727:509--532.

\bibitem{Kunnen2016}
Kunnen~RPJ, Ostilla-M{\'o}nico~R, van~der Poel~EP, et~al. Transition to
  geostrophic convection: the role of the boundary conditions. J Fluid Mech.
  2016;\hspace{0pt}799:413--432.

\bibitem{Stellmach2014}
Stellmach~S, Lischper~M, Julien~K, et~al. {Approaching the asymptotic regime of
  rapidly rotating convection: Boundary layers versus interior dynamics}. Phys
  Rev Lett. 2014;\hspace{0pt}113(25):254501.

\bibitem{Stevens2010b}
Stevens~RJAM, Clercx~HJH, Lohse~D. {Optimal Prandtl number for heat transfer in
  rotating {Rayleigh}--B{\'e}nard convection}. New J Phys.
  2010;\hspace{0pt}12(7):075005.

\bibitem{Stevens2011}
Stevens~RJAM, Lohse~D, Verzicco~R. {Prandtl and {Rayleigh} number dependence of
  heat transport in high {Rayleigh} number thermal convection}. J Fluid Mech.
  2011;\hspace{0pt}688(1):31--43.

\bibitem{Stevens2013}
Stevens~RJAM, Clercx~HJH, Lohse~D. {Heat transport and flow structure in
  rotating {Rayleigh}--B{\'e}nard convection}. Eur J Mech (B/Fluids).
  2013;\hspace{0pt}40:41--49.

\bibitem{Weiss2010}
Weiss~S, Stevens~RJAM, Zhong~JQ, et~al. {Finite-Size Effects Lead to
  Supercritical Bifurcations in Turbulent Rotating {Rayleigh}--B\'enard
  Convection}. Phys Rev Lett. 2010 Nov;\hspace{0pt}105:224501.

\bibitem{Weiss2011a}
Weiss~S, Ahlers~G. {Heat transport by turbulent rotating {Rayleigh}--B{\'e}nard
  convection and its dependence on the aspect ratio}. J Fluid Mech.
  2011;\hspace{0pt}684(407):205.

\bibitem{Zhong2009}
Zhong~JQ, Stevens~RJAM, Clercx~HJH, et~al. {Prandtl-, {Rayleigh}-, and
  Rossby-Number dependence of heat transport in turbulent rotating
  {Rayleigh}--B\'enard Convection}. Phys Rev Lett. 2009
  Jan;\hspace{0pt}102(4):044502.

\bibitem{Oberbeck1879}
Oberbeck~A. {Ueber die W\"armeleitung der Fl\"ussigkeiten bei
  Ber\"ucksichtigung der Str\"omungen infolge von Temperaturdifferenzen}.
  Annalen der Physik. 1879;\hspace{0pt}243(6):271--292.

\bibitem{Boussinesq1903}
Boussinesq~JV. {Th{\'e}orie analytique de la chaleur}. Vol.~2. Gauthier-Villars
  Paris; 1903.

\bibitem{Becker2006}
Becker~N, Scheel~JD, Cross~MC, et~al. {Effect of the centrifugal force on
  domain chaos in Rayleigh-B{\'e}nard convection}. Phys Rev E.
  2006;\hspace{0pt}73(6):066309.

\bibitem{Barcilon1967}
Barcilon~V, Pedlosky~J. On the steady motions produced by a stable
  stratification in a rapidly rotating fluid. J Fluid Mech.
  1967;\hspace{0pt}29(04):673--690.

\bibitem{Brummell2000}
Brummell~N, Hart~JE, Lopez~JM. On the flow induced by centrifugal buoyancy in a
  differentially-heated rotating cylinder. Theoret Comput Fluid Dynamics.
  2000;\hspace{0pt}14(1):39--54.

\bibitem{Curbelo2014}
Curbelo~J, Lopez~JM, Mancho~AM, et~al. {Confined rotating convection with large
  Prandtl number: Centrifugal effects on wall modes}. Phys Rev E.
  2014;\hspace{0pt}89(1):013019.

\bibitem{Hart1999}
Hart~J, Ohlsen~D. On the thermal offset in turbulent rotating convection. Phys
  Fluids. 1999;\hspace{0pt}11:2101.

\bibitem{Hart2000}
Hart~JE. On the influence of centrifugal buoyancy on rotating convection. J
  Fluid Mech. 2000;\hspace{0pt}403:133--151.

\bibitem{Homsy1969}
Homsy~GM, Hudson~JL. Centrifugally driven thermal convection in a rotating
  cylinder. J Fluid Mech. 1969;\hspace{0pt}35(1):33--52.

\bibitem{Homsy1971b}
Homsy~GM, Hudson~JL. Centrifugal convection and its effect on the asymptotic
  stability of a bounded rotating fluid heated from below. J Fluid Mech.
  1971;\hspace{0pt}48(03):605--624.

\bibitem{Lopez2006}
Lopez~JM, Rubio~A, Marques~F. Travelling circular waves in axisymmetric
  rotating convection. J Fluid Mech. 2006;\hspace{0pt}569:331--348.

\bibitem{Lopez2009}
Lopez~J, Marques~F. Centrifugal effects in rotating convection: nonlinear
  dynamics. J Fluid Mech. 2009;\hspace{0pt}628:269--297.

\bibitem{Marques2007}
Marques~F, Mercader~I, Batiste~O, et~al. Centrifugal effects in rotating
  convection: axisymmetric states and three-dimensional instabilities. J Fluid
  Mech. 2007;\hspace{0pt}580:303.

\bibitem{Torrest1974}
Torrest~MA, Hudson~JL. The effect of centrifugal convection on the stability of
  a rotating fluid heated from below. Appl Sci Res.
  1974;\hspace{0pt}29(1):273--289.

\bibitem{Dowell2005}
Dowell~DC, Alexander~CR, Wurman~JM, et~al. {Centrifuging of hydrometeors and
  debris in tornadoes: Radar-reflectivity patterns and wind-measurement
  errors}. Mon Wea Rev. 2005;\hspace{0pt}133(6):1501--1524.

\bibitem{Orf2017}
Orf~L, Wilhelmson~R, Lee~B, et~al. Evolution of a long-track violent tornado
  within a simulated supercell. Bull Amer Meteor Soc.
  2017;\hspace{0pt}98(1):45--68.

\bibitem{Goliger1998}
Goliger~AM, Milford~R. A review of worldwide occurrence of tornadoes. J Wind
  Eng Ind Aerodyn. 1998;\hspace{0pt}74:111--121.

\bibitem{Balme2006}
Balme~M, Greeley~R. {Dust devils on Earth and Mars}. Rev Geophys.
  2006;\hspace{0pt}44(3).

\bibitem{Thomas1985}
Thomas~P, Gierasch~PJ. {Dust devils on Mars}. Science.
  1985;\hspace{0pt}230(4722):175--177.

\bibitem{Davies-Jones2015}
Davies-Jones~R. A review of supercell and tornado dynamics. Atmos Res.
  2015;\hspace{0pt}158:274--291.

\bibitem{Klemp1987}
Klemp~JB. Dynamics of tornadic thunderstorms. Ann Rev Fluid Mech.
  1987;\hspace{0pt}19(1):369--402.

\bibitem{Lemon1979}
Lemon~LR, Doswell~III~CA. Severe thunderstorm evolution and mesocyclone
  structure as related to tornadogenesis. Mon Wea Rev.
  1979;\hspace{0pt}107(9):1184--1197.

\bibitem{Bluestein2013}
Bluestein~HB. Severe convective storms and tornadoes. Springer; 2013.

\bibitem{Lin1992}
Lin~SJ. Contour dynamics of tornado-like vortices. J Atmos Sci.
  1992;\hspace{0pt}49(18):1745--1756.

\bibitem{Snow1987}
Snow~JT. Atmospheric columnar vortices. Rev Geophys.
  1987;\hspace{0pt}25(3):371--385.

\bibitem{Maxworthy1973}
Maxworthy~T. A vorticity source for large-scale dust devils and other comments
  on naturally occurring columnar vortices. J Atmos Sci.
  1973;\hspace{0pt}30(8):1717--1722.

\bibitem{Wurman2012}
Wurman~J, Dowell~D, Richardson~Y, et~al. The second verification of the origins
  of rotation in tornadoes experiment: Vortex2. Bull Amer Meteor Soc.
  2012;\hspace{0pt}93(8):1147--1170.

\bibitem{Markowski2014}
Markowski~P, Richardson~Y. What we know and don't know about tornado formation.
  Phys Today. 2014;\hspace{0pt}67(9):26.

\bibitem{Trapp2005}
Trapp~RJ, Stumpf~GJ, Manross~KL. A reassessment of the percentage of tornadic
  mesocyclones. Wea Forecasting. 2005;\hspace{0pt}20(4):680--687.

\bibitem{Marquis2012}
Marquis~J, Richardson~Y, Markowski~P, et~al. {Tornado maintenance investigated
  with high-resolution dual-Doppler and EnKF analysis}. Mon Wea Rev.
  2012;\hspace{0pt}140(1):3--27.

\bibitem{Rasmussen1994}
Rasmussen~EN, Straka~JM, Davies-Jones~R, et~al. {Verification of the origins of
  rotation in tornadoes experiment: VORTEX}. Bull Amer Meteor Soc.
  1994;\hspace{0pt}75(6):995--1006.

\bibitem{Fiedler1995}
Fiedler~BH. On modelling tornadoes in isolation from the parent storm.
  Atmos-Ocean. 1995;\hspace{0pt}33(3):501--512.

\bibitem{Ward1972}
Ward~NB. The exploration of certain features of tornado dynamics using a
  laboratory model. J Atmos Sci. 1972;\hspace{0pt}29(6):1194--1204.

\bibitem{Houser2018}
Houser~J, Bluestein~H, Seimon~A, et~al. Rapid-scan mobile radar observations of
  tornadogenesis. In: AGU Fall Meeting Abstracts; 2018.

\bibitem{Rotunno2013}
Rotunno~R. The fluid dynamics of tornadoes. Ann Rev Fluid Mech.
  2013;\hspace{0pt}45.

\bibitem{Leslie1971}
Leslie~L. {The development of concentrated vortices: A numerical study}. J
  Fluid Mech. 1971;\hspace{0pt}48(1):1--21.

\bibitem{Smith1978}
Smith~RK, Leslie~LM. Tornadogenesis. Quart J Roy Meteor Soc.
  1978;\hspace{0pt}104(439):189--198.

\bibitem{Trapp1997}
Trapp~RJ, Davies-Jones~R. Tornadogenesis with and without a dynamic pipe
  effect. J Atm Sci. 1997;\hspace{0pt}54(1):113--133.

\bibitem{Trapp1999}
Trapp~R, Mitchell~E, Tipton~G, et~al. {Descending and nondescending tornadic
  vortex signatures detected by WSR-88Ds}. Wea Forecasting.
  1999;\hspace{0pt}14(5):625--639.

\bibitem{Houser2015}
Houser~JL, Bluestein~HB, Snyder~JC. {Rapid-scan, polarimetric, Doppler radar
  observations of tornadogenesis and tornado dissipation in a tornadic
  supercell: The “El Reno, Oklahoma” storm of 24 May 2011}. Mon Wea Rev.
  2015;\hspace{0pt}143(7):2685--2710.

\bibitem{French2013}
French~MM, Bluestein~HB, PopStefanija~I, et~al. Reexamining the vertical
  development of tornadic vortex signatures in supercells. Mon Wea Rev.
  2013;\hspace{0pt}141(12):4576--4601.

\bibitem{Bluestein2019}
Bluestein~HB, Thiem~KJ, Snyder~JC, et~al. {Tornadogenesis and early tornado
  evolution in the El Reno, Oklahoma, supercell on 31 May 2013}. Mon Wea Rev.
  2019;\hspace{0pt}.

\bibitem{Lewellen2000}
Lewellen~DC, Lewellen~WS, Xia~J. The influence of a local swirl ratio on
  tornado intensification near the surface. J Atmos Sci.
  2000;\hspace{0pt}57(4):527--544.

\bibitem{Church1979}
Church~C, Snow~J, Baker~G, et~al. {Characteristics of tornado-like vortices as
  a function of swirl ratio: A laboratory investigation}. J Atm Sci.
  1979;\hspace{0pt}36(9):1755--1776.

\bibitem{Fiedler1994}
Fiedler~BH. The thermodynamic speed limit and its violation in axisymmetric
  numerical simulations of tornado-like vortices. Atmos-Ocean.
  1994;\hspace{0pt}32(2):335--359.

\bibitem{Fiedler1998}
Fiedler~BH. Wind-speed limits in numerically simulated tornadoes with suction
  vortices. Quart J Roy Meteor Soc. 1998;\hspace{0pt}124(551):2377--2392.

\bibitem{Fiedler2009}
Fiedler~B. Suction vortices and spiral breakdown in numerical simulations of
  tornado-like vortices. Atmos Sci Lett. 2009;\hspace{0pt}10(2):109--114.

\bibitem{Nolan1999}
Nolan~DS, Farrell~BF. The structure and dynamics of tornado-like vortices. J
  Atmos Sci. 1999;\hspace{0pt}56(16):2908--2936.

\bibitem{Nolan2012}
Nolan~DS. Three-dimensional instabilities in tornado-like vortices with
  secondary circulations. J Fluid Mech. 2012;\hspace{0pt}711:61--100.

\bibitem{Nolan2005}
Nolan~DS. A new scaling for tornado-like vortices. J Atmos Sci.
  2005;\hspace{0pt}62(7):2639--2645.

\bibitem{Rotunno2016}
Rotunno~R, Bryan~GH, Nolan~DS, et~al. {Axisymmetric tornado simulations at high
  Reynolds number}. J Atmos Sci. 2016;\hspace{0pt}73(10):3843--3854.

\bibitem{Vogt2013}
Vogt~T, Grants~I, Eckert~S, et~al. Spin-up of a magnetically driven
  tornado-like vortex. J Fluid Mech. 2013;\hspace{0pt}736:641--662.

\bibitem{Castano2017}
Castano~D, Navarro~M, Herrero~H. Double vortices and single-eyed vortices in a
  rotating cylinder under thermal gradients. Computers \& Mathematics with
  Applications. 2017;\hspace{0pt}73(10):2238--2257.

\bibitem{Lewellen1993}
Lewellen~W. Tornado vortex theory. Washington DC AGU Geophys Monograph Series.
  1993;\hspace{0pt}79:19--39.

\bibitem{Ahlers2006}
Ahlers~G, Brown~E, {Fontenele Araujo}~F, et~al. {Non-Oberbeck--Boussinesq
  effects in strongly turbulent {Rayleigh}--B{\'e}nard convection}. J\ Fluid
  Mech. 2006;\hspace{0pt}569:409--445.

\bibitem{Gray1976}
Gray~DD, Giorgini~A. {The validity of the Boussinesq approximation for liquids
  and gases}. Int J Heat Mass Transfer. 1976;\hspace{0pt}19:545--551.

\bibitem{Horn2013a}
Horn~S, Shishkina~O, Wagner~C. {On non-Oberbeck--Boussinesq effects in
  three-dimensional {Rayleigh}--B{\'e}nard convection in glycerol}. J Fluid
  Mech. 2013;\hspace{0pt}724:175--202.

\bibitem{Sugiyama2009}
Sugiyama~K, Calzavarini~E, Grossmann~S, et~al. {Flow organization in
  two-dimensional non-Oberbeck--Boussinesq {Rayleigh}-B{\'e}nard convection in
  water}. J Fluid Mech. 2009;\hspace{0pt}637:105--135.

\bibitem{Shishkina2015}
Shishkina~O, Horn~S, Wagner~S, et~al. {Thermal boundary layer equation for
  turbulent Rayleigh--B{\'e}nard convection}. Phys Rev Lett.
  2015;\hspace{0pt}114(11):114302.

\bibitem{Shishkina2016}
Shishkina~O, Horn~S. Thermal convection in inclined cylindrical containers. J
  Fluid Mech. 2016;\hspace{0pt}790:R3.

\bibitem{Shishkina2010}
Shishkina~O, Stevens~RJAM, Grossmann~S, et~al. Boundary layer structure in
  turbulent thermal convection and its consequences for the required numerical
  resolution. New J Phys. 2010;\hspace{0pt}12(7):075022.

\bibitem{Davies-Jones2003}
Davies-Jones~R. An expression for effective buoyancy in surroundings with
  horizontal density gradients. J Atm Sci. 2003;\hspace{0pt}60(23):2922--2925.

\bibitem{Doswell2004}
Doswell~III~CA, Markowski~PM. Is buoyancy a relative quantity? Mon Wea Rev.
  2004;\hspace{0pt}132(4):853--863.

\bibitem{Wang2020}
Wang~BF, Zhou~Q, Sun~C. Vibration-induced boundary-layer destabilization
  achieves massive heat-transport enhancement. Science advances.
  2020;\hspace{0pt}6(21):eaaz8239.

\bibitem{Cheng2016}
Cheng~JS, Aurnou~JM. Tests of diffusion-free scaling behaviors in numerical
  dynamo datasets. Earth Planet Sci Lett. 2016;\hspace{0pt}436:121--129.

\bibitem{Aurnou2020}
Aurnou~JM, Horn~S, Julien~K. Connections between nonrotating, slowly rotating,
  and rapidly rotating turbulent convection transport scalings. Physical Review
  Research. 2020;\hspace{0pt}2(4):043115.

\bibitem{Ahlers2009}
Ahlers~G, Grossmann~S, Lohse~D. {Heat transfer and large scale dynamics in
  turbulent {Rayleigh}-B{\'e}nard convection}. Rev Mod Phys.
  2009;\hspace{0pt}81(2):503.

\bibitem{Grossmann2000}
Grossmann~S, Lohse~D. Scaling in thermal convection: A unifying theory. J Fluid
  Mech. 2000;\hspace{0pt}407:27--56.

\bibitem{Julien2012b}
Julien~K, Rubio~AM, Grooms~I, et~al. Statistical and physical balances in low
  rossby number rayleigh--b{\'e}nard convection. Geophys \& Astrophys Fluid
  Dyn. 2012;\hspace{0pt}106(4-5):392--428.

\bibitem{King2012b}
King~EM, Stellmach~S, Aurnou~JM. {Heat transfer by rapidly rotating
  {Rayleigh}--B{\'e}nard convection}. J Fluid Mech.
  2012;\hspace{0pt}691:568--582.

\bibitem{Stevens2009}
Stevens~RJAM, Zhong~JQ, Clercx~HJH, et~al. {Transitions between Turbulent
  States in Rotating {Rayleigh}--B\'enard Convection}. Phys Rev Lett. 2009
  Jul;\hspace{0pt}103:024503.

\bibitem{Kunnen2011}
Kunnen~RPJ, Stevens~RJAM, Overkamp~J, et~al. {The role of Stewartson and Ekman
  layers in turbulent rotating {Rayleigh}--B\'enard convection}. J Fluid Mech.
  2011;\hspace{0pt}688:422--442.

\bibitem{Willoughby1990}
Willoughby~HE. Gradient balance in tropical cyclones. J Atmos Sci.
  1990;\hspace{0pt}47(2):265--274.

\bibitem{Davies-Jones1973}
Davies-Jones~RP. The dependence of core radius on swirl ratio in a tornado
  simulator. J Atmos Sci. 1973;\hspace{0pt}30(7):1427--1430.

\bibitem{Cheng2015}
Cheng~JS, Stellmach~S, Ribeiro~A, et~al. {Laboratory-numerical models of
  rapidly rotating convection in planetary cores}. Geophys J Int.
  2015;\hspace{0pt}201(1):1--17.

\bibitem{Sprague2006}
Sprague~M, Julien~K, Knobloch~E, et~al. Numerical simulation of an
  asymptotically reduced system for rotationally constrained convection. J
  Fluid Mech. 2006 Mar;\hspace{0pt}551:141--174.

\bibitem{Grooms2010}
Grooms~I, Julien~K, Weiss~JB, et~al. {Model of Convective {Taylor} Columns in
  Rotating {Rayleigh}-B\'enard Convection}. Phys Rev Lett. 2010
  Jun;\hspace{0pt}104:224501.

\bibitem{Snow1984}
Snow~JT. On the formation of particle sheaths in columnar vortices. J Atm Sci.
  1984;\hspace{0pt}41(16):2477--2491.

\bibitem{Wurman1996}
Wurman~J, Straka~JM, Rasmussen~EN. {Fine-scale Doppler radar observations of
  tornadoes}. Science. 1996;\hspace{0pt}272(5269):1774--1777.

\bibitem{Fiedler1986}
Fiedler~BH, Rotunno~R. A theory for the maximum windspeeds in tornado-like
  vortices. J Atmos Sci. 1986;\hspace{0pt}43(21):2328--2340.

\bibitem{Brandes1984}
Brandes~EA. Relationships between radar-derived thermodynamic variables and
  tornadogenesis. Mon Wea Rev. 1984;\hspace{0pt}112(5):1033--1052.

\bibitem{Dowell2002a}
Dowell~DC, Bluestein~HB. {The 8 June 1995 McLean, Texas, storm. Part I:
  Observations of cyclic tornadogenesis}. Mon Wea Rev.
  2002;\hspace{0pt}130(11):2626--2648.

\bibitem{Oruba2017}
Oruba~L, Davidson~P, Dormy~E. Eye formation in rotating convection. J Fluid
  Mech. 2017;\hspace{0pt}812:890--904.

\bibitem{Tanamachi2007}
Tanamachi~RL, Bluestein~HB, Lee~WC, et~al. Ground-based velocity track display
  (gbvtd) analysis of w-band doppler radar data in a tornado near stockton,
  kansas, on 15 may 1999. Monthly weather review.
  2007;\hspace{0pt}135(3):783--800.

\bibitem{Bluestein1993}
Bluestein~HB, Ladue~JG, Stein~H, et~al. Doppler radar wind spectra of supercell
  tornadoes. Mon Wea Rev. 1993;\hspace{0pt}121(8):2200--2222.

\bibitem{Greenspan1968}
Greenspan~HP. The theory of rotating fluids. Cambridge University Press
  (London); 1968.

\bibitem{Fiedler1989}
Fiedler~BH. Conditions for laminar flow in geophysical vortices. J Atm Sci.
  1989;\hspace{0pt}46(2):252--260.

\bibitem{Lilly1986b}
Lilly~DK. The structure, energetics and propagation of rotating convective
  storms. part ii: Helicity and storm stabilization. J Atmos Sci.
  1986;\hspace{0pt}43(2):126--140.

\bibitem{Sullivan1959}
Sullivan~RD. A two-cell vortex solution of the navier-stokes equations. Journal
  of the Aerospace Sciences. 1959;\hspace{0pt}26(11):767--768.

\bibitem{Fujita1985}
Fujita~TT, Stiegler~D. {Detailed Analysis of the Tornado Outbreak in the
  Carolinas by Using Radar, Satellite, and Aerial Survey Data}. Preprints, 14th
  Conf on Severe Local Storms. 1985;\hspace{0pt}:271--274.

\bibitem{Cushman2011}
Cushman-Roisin~B, Beckers~JM. Introduction to geophysical fluid dynamics:
  physical and numerical aspects. Vol. 101. Academic Press; 2011.

\bibitem{Spiegel1971}
Spiegel~EA. {Convection in stars I. Basic Boussinesq convection}. Ann Rev
  Astron Astrophys. 1971;\hspace{0pt}9(1):323--352.

\bibitem{Batchelor1967}
Batchelor~G. An introduction to fluid dynamics. Cambridge University Press;
  1967.

\bibitem{Dowell2002b}
Dowell~DC, Bluestein~HB. {The 8 June 1995 McLean, Texas, storm. Part II: Cyclic
  tornado formation, maintenance, and dissipation}. Mon Wea Rev.
  2002;\hspace{0pt}130(11):2649--2670.

\end{thebibliography}

\end{document}